\newif\ifconfver
\confvertrue       

\ifconfver
    \documentclass[10pt,twocolumn,twoside]{IEEEtran}
\else
    \documentclass[11pt,draftcls,onecolumn,twoside]{IEEEtran}
\fi

\usepackage{amsmath,amsfonts,amssymb,amsbsy,bm,paralist,theorem,ifthen,color}
\usepackage{calc}
\usepackage{caption,graphicx,subfig,stfloats}
\usepackage[numbers,sort&compress]{natbib}
\usepackage{hyperref}
\usepackage{algorithmic,algorithm}
\hypersetup{unicode,bookmarksnumbered,bookmarksopen=true,bookmarksopenlevel=2,
            breaklinks=true,colorlinks=true,linkcolor=blue,citecolor=blue,filecolor=black,urlcolor=blue,
            plainpages=false,pdfstartview=FitH,
            pdfsubject = WIET in MISO-IC,
            pdfauthor = Chao Shen and Wei-Chiang Li and Tsung-Hui Chang,
            pdftitle = Wireless Information and Energy Transfer in Multi-Antenna Interference Channel,
            pdfkeywords = {wireless energy transfer, energy harvesting, interference channel, beamforming, convex optimization}
           }

\newcommand\hb{\ensuremath{{\bm h}}}
\newcommand\xb{\ensuremath{{\bm x}}}

\newcommand\vb{\ensuremath{{\bm v}}}
\newcommand\Ab{\ensuremath{{\bm A}}}
\newcommand\Hb{\ensuremath{{\bm H}}}

\newcommand\Sb{\ensuremath{{\bm S}}}
\newcommand\Xb{\ensuremath{{\bm X}}}
\newcommand\Qb{\ensuremath{{\bm Q}}}

\newcommand\Wb{\ensuremath{{\bm W}}}
\newcommand\zerob{\ensuremath{{\bm 0}}}
\newcommand\Psib {\ensuremath{{\bm\Psi}}}
\newcommand{\E}         {\ensuremath{{\mathbb{E}}}}
\newcommand{\rank}      {\ensuremath{{\rm rank}}}
\newcommand{\SNR}       {\ensuremath{{\rm SNR}}}
\newcommand{\range}     {\ensuremath{{\rm Range}}}
\newcommand{\st}        {\ensuremath{{\rm s.t.}}}
\newcommand{\tr}        {\ensuremath{{\rm Tr}}}
\newcommand{\Tr}        {\ensuremath{{\rm Tr}}}
\newcommand{\CN}        {\ensuremath{\mathcal{CN}}}
\newcommand{\Rset}      {\ensuremath{\mathbb{R}}}
\newcommand{\Cset}      {\ensuremath{\mathbb{C}}}
\DeclareMathOperator*   {\argmax}{arg\,max}
\newtheorem{Lemma}{Lemma}
\newtheorem{Proposition}{Proposition}
\newtheorem{Property}{Property}
\newtheorem{Remark}{Remark}

\graphicspath{{fig/}}

\begin{document}
\bibliographystyle{IEEEtran}
\title{Wireless Information and Energy Transfer in Multi-Antenna Interference Channel}

\ifconfver \else {\linespread{1.1} \rm \fi

\author{\vspace{0.8cm}Chao Shen$^\star$, Wei-Chiang Li$^\dag$ and Tsung-Hui Chang$^{\ddag}$ \\
\thanks{
The work of Chao Shen is supported by
the Opening Project of The State Key Laboratory of Integrated Services Networks, Xidian University (Grant No. ISN14-09),
the China Postdoctoral Science Foundation (Grant No. 2013M530519),
the National Natural Science Foundation of China (Grant No. 61222105),
the Key Project of State Key Lab of Rail Traffic and Control (Grant No. RCS2012ZZ004), Beijing Jiaotong University,
the Key grant Project of Chinese Ministry of Education (No. 313006),
and the Fundamental Research Funds for the Central Universities (Grant No. 2010JBZ008 and 2012YJS017).
The work of Tsung-Hui Chang is supported by National Science Council, Taiwan (R.O.C.), by Grant NSC 102-2221-E-011-005-MY3.
Part of this work was presented in IEEE GLOBECOM 2012 \cite{ChaoGlobecom2012}.}
\thanks{$^\star$Chao Shen is with
the State Key Laboratory of Rail Traffic Control and Safety, Beijing Jiaotong University, Beijing, China and the State Key Laboratory of Integrated Services Networks, Xidian University, Xi'an, China. E-mail: chaoshen@ieee.org.}
\thanks{$^\dag$Wei-Chiang Li is with Institute of Communications Engineering, National Tsing Hua University, Hsinchu 30013, Taiwan, (R.O.C.). E-mail: weichiangli@gmail.com.}
\thanks{$^{\ddag}$Tsung-Hui Chang is the corresponding author. Address:
Department of Electronic and Computer Engineering, National Taiwan University of Science and Technology, Taipei 10607, Taiwan, (R.O.C.). E-mail: tsunghui.chang@ieee.org.}
}

\markboth{Submitted to IEEE TRANSACTIONS ON SIGNAL PROCESSING}{Submitted to IEEE TRANSACTIONS ON SIGNAL PROCESSING} \maketitle

\begin{abstract}
This paper considers the transmitter design for wireless information and energy transfer (WIET) in a multiple-input single-output (MISO) interference channel (IFC). The design problem is to maximize the system throughput (i.e., the weighted sum rate) subject to individual energy harvesting constraints and power constraints. Different from the conventional IFCs without energy harvesting, the cross-link signals in the considered scenario play two opposite roles in information detection (ID) and energy harvesting (EH). It is observed that the ideal scheme, where the receivers can simultaneously perform ID and EH from the received signal, may not always achieve the best tradeoff between information transfer and energy harvesting, but simple practical schemes based on time splitting may perform better. We therefore propose two practical time splitting schemes, namely time division mode switching (TDMS) and time division multiple access (TDMA), in addition to a power splitting (PS) scheme which separates the received signal into two parts for ID and EH, respectively.
In the two-user scenario, we show that beamforming is optimal to all the schemes. Moreover, the design problems associated with the TDMS and TDMA schemes admit semi-analytical solutions.
In the general $K$-user scenario, a successive convex approximation method is proposed to handle the WIET problems associated with the ideal scheme and the PS scheme, which are known to be NP-hard in general. The $K$-user TDMS and TDMA schemes are shown efficiently solvable as convex problems. Simulation results show that stronger cross-link channel powers actually improve the information sum rate under energy harvesting constraints. Moreover, none of the schemes under consideration can dominate another in terms of the sum rate performance.
\noindent {\bfseries EDICS}:  SPC-APPL, SPC-INTF, SPC-CCMC, SAM-BEAM
\end{abstract}

\ifconfver \else \IEEEpeerreviewmaketitle} \fi
\ifconfver \else \newpage \fi

\section{Introduction}\label{sec: intro}
Recently, scavenging energy from the environment has been considered as a potential approach to
prolonging the lifetime of battery-powered sensor networks and to implementing self-sustained communication systems. For example, the base stations may be powered by wind mills or solar photovoltaic (PV) arrays, and can harvest energy for providing services to the mobile users. This idea has motivated considerable research endeavors in the past few years, investigating wireless systems with energy-harvesting transmitters; see, e.g., \cite{Ozel2011,Xu2012,Huang_Zhang_Cui2012,Tutuncuoglu2012_JCN,Lee2012}. In these works, optimal transmission strategies under energy-harvesting constraints are studied from single-input single-output (SISO) channels to complex interference channels (IFCs). In contrast to the base stations, it may be difficult for the mobile devices and sensor nodes to harvest energy from the sun and wind effectively. One possible solution to this issue is \emph{wireless energy transfer (WET)}, that is, the power-connected transmitters transfer energy wirelessly to charge the mobile devices. A successful application of WET is the radio frequency identification (RFID) system where the receiver wirelessly charges energy from the transmitter (through induction coupling) and use the energy to communicate with the transmitter. The works in \cite{Kurs2007,Karalis2008} showed that, using coupled magnetic resonances, energy can be wirelessly transferred for two meters with over $50\%$ energy conversion efficiency. WET can also be achieved via the RF electromagnetic signals; see \cite{Dolgov2010,Nintanavongsa2012} for recent developments of RF-based energy harvesting circuits. Compared to the techniques based on induction and magnetic resonance coupling, RF signals can achieve long-distance WET; however, the energy conversion efficiency is in general low. This calls for advanced signal processing techniques, such as beamforming, to improve the energy conversion efficiency.

Since the RF signals can carry both information and energy, in recent years, it has been of great interest to study wireless communication systems where the receivers can not only decode information bits but also harvest energy from the received RF signals, i.e., \textit{wireless information and energy transfer} (WIET) systems \cite{Varshney2008,Grover2010,Fouladgar2012,Zhou2012,Zhou2012_jrnl,Zhangrui2011,Shi2013}.
Specifically, in \cite{Varshney2008}, the optimal tradeoff between information capacity and energy transfer of the WIET system was studied for a SISO flat fading channel. In \cite{Grover2010}, the optimal power allocation strategy for a SISO frequency-selective fading channel was derived under a receiver energy harvesting constraint. The work in \cite{Fouladgar2012} further extends these studies to the
multiple access channel (MAC) and two-hop relay network with an energy harvesting relay. It was shown that in general there exist nontrivial tradeoffs between information transfer and energy harvesting. The works in \cite{Varshney2008,Grover2010,Fouladgar2012} assume the \emph{ideal receivers} which can decode information bits and harvest energy from the received RF signals \emph{simultaneously}. Unfortunately, current circuit technologies cannot achieve this yet. In view of this, practical WIET schemes are proposed. In particular, Zhou et al. proposed in \cite{Zhou2012,Zhou2012_jrnl} a dynamic power splitting (PS) scheme for a SISO flat fading channel, wherein, the received RF signal is either used for information detection (ID), energy harvesting (EH), or is split into two parts, one for ID and the other for EH. Considering a multiple-input multiple-output (MIMO) flat-fading channel, in addition to the PS scheme, the authors in \cite{Zhangrui2011} further proposed a time switching scheme where the receiver performs ID in one time slot while EH in the other time slot. In \cite{Shi2013}, the dynamic PS scheme was extended to a multi-user multiple-input single-output (MISO) broadcast channel, and the optimal transmit beamforming and power splitting coefficients are jointly optimized to minimize the transmission power subject to information rate and energy harvesting constraints.

In this paper, we consider a $K$-user MISO interference channel and study the optimal transmission strategies for WIET. We first consider the ideal receivers, and formulate the design problem as a weighted sum rate maximization problem subject to individual energy harvesting constraints and power constraints. It is interesting to note that, different from the conventional IFCs without energy harvesting, the cross-link signals in the considered scenario can degrade the information sum rate on one hand, but, at the same time, boost energy harvesting of the receivers on the other hand. And it 
turns out that the ideal scheme with ideal receivers may not always perform best in the complex interference environment, but simple practical schemes based on time splitting may instead yield better sum rate performance. This is in sharp contrast to the scenarios studied in \cite{Zhou2012,Zhou2012_jrnl,Zhangrui2011,Shi2013} where time splitting schemes usually exhibit poorer performance. This intriguing observation motivates us to propose two practical WIET schemes for the MISO IFC, namely, the time division mode switching (TDMS) scheme and the time division multiple access (TDMA) scheme\footnote{As will be shown in Section \ref{subsec:TDMA_A}, the proposed TDMA scheme is similar to but not completely the same as the TDMA scheme in conventional IFCs without energy harvesting.}, in addition to the PS scheme \cite{Zhou2012_jrnl}. In the TDMS scheme, the transmission time is divided into two time slots. All receivers perform EH in the first time slot and subsequently perform ID in the second time slot. The TDMA scheme divides the transmission time into $K$ time slots, and in each time slot, one receiver performs ID while the others perform EH. We analytically show how the design problems associated with the three schemes can be efficiently handled. Specifically, for the two-user scenario, we show that transmit beamforming is an optimal transmission strategy for all schemes. Moreover, the design problems associated with the TDMS and TDMA schemes admit semi-analytical solutions in the two-user scenario and can be solved as convex problems in the general $K$-user scenario. Since the WIET design problems associated with the ideal scheme and the PS scheme in the $K$-user scenario are NP-hard in general, we further present an efficient approximation method based on the log-exponential reformulation and successive convex approximation techniques \cite{WCLi2013}.
The presented simulation results will show that stronger cross-link channel powers actually improve the information sum rate under energy harvesting constraints. Moreover, the three schemes do not dominate each other in terms of sum rate performance. Roughly speaking, if the cross-link channel powers are not strong or the energy harvesting constraints are not stringent, the PS scheme can outperform TDMS and TDMA schemes; otherwise, the TDMS scheme can perform best. In some interference dominated scenarios, the TDMS scheme and TDMA scheme even outperform the ideal scheme.

The rest of this paper is organized as follows. In Section \ref{sec:sig_mod_prob_formu}, the signal model of the MISO interference channel is presented. Starting with the two-user scenario, in Section \ref{sec:ideal receiver}, the optimal WIET transmission strategy for ideal receivers is analyzed. The result motivates the developments of the practical TDMS and TDMA schemes, which are presented in Section \ref{sec:prac_trans}. Section \ref{sec: k user} extends the study to the general $K$-user scenario; the design problem of the PS scheme is also presented in that section. Simulation results are presented in Section \ref{sec: simulation}. The conclusions and discussion of future researches are given in Section \ref{sec: conclusions}.

\textit{Notations:} Column vectors and matrices are written in boldfaced lowercase and uppercase letters, e.g., ${\bm{a}}$ and $\Ab$. The superscripts $(\cdot)^T$, $(\cdot)^H$ and $(\cdot)^{-1}$ represent the transpose, (Hermitian) conjugate transpose and matrix inverse, respectively. ${\rm rank}(\Ab)$ and $\Tr(\Ab)$ represent the rank and trace of matrix $\Ab$, respectively. $\Ab\succeq \zerob$ ($\succ \zerob$) means that matrix $\Ab$ is positive semidefinite (positive definite). $\|{\bm{a}}\|$ denotes the Euclidean norm of vector ${\bm{a}}$. The orthogonal projection onto the column space of a tall matrix $\Ab$ is denoted by $\Pi_\Ab\triangleq\Ab(\Ab^H\Ab)^{-1}\Ab^H$. Moreover, the projection onto the orthogonal complement of the column space of $\Ab$ is denoted by $\Pi_\Ab^\perp\triangleq\mathbf{I}-\Pi_\Ab$ where $\mathbf{I}$ is the identity matrix.

\vspace{-3mm}
\section{Signal Model and Problem Statement}\label{sec:sig_mod_prob_formu}
We consider a multi-user interference channel with $K$ pairs of transmitters and receivers communicating over a common frequency band.
Each of the transmitters is equipped with $N_t$ antennae, while each of the receivers has single antenna.
Let $\xb_i \in \mathbb{C}^{N_t}$ be the signal vector transmitted by transmitter $i$, and $\hb_{ik} \in \mathbb{C}^{N_t}$ be the channel vector from transmitter $i$ to receiver $k$, for all $i,k\in\{1,2,\ldots,K\}$. The received signal at receiver $i$ is given by
\begin{align}\label{eq: received signal}
    y_i = \hb_{ii}^H\xb_i+\sum_{k=1,k\neq i}^K\hb_{ki}^H\xb_k+n_i,~i=1,\ldots,K,
\end{align}
where $n_i\sim\CN(0,\sigma_i^2)$ is the additive Gaussian noise at receiver $i$.
Unlike the conventional MISO IFC \cite{Jorswieck08} where the receivers focus only on extracting information, we consider in this paper that the receivers can also scavenge energy from the received signals \cite{Varshney2008,Grover2010,Zhangrui2011}, i.e, energy harvesting. Therefore, in addition to information, the transmitters can also wirelessly transfer energy to the receivers. We call the two operation modes the information detection (ID) mode and the energy harvesting (EH) mode, respectively.

Assume that $\xb_i$ contains the information intended for receiver $i$ which is Gaussian encoded with zero mean and covariance matrix $\Sb_i\succeq \zerob$, i.e., $\xb_i\sim\CN(\zerob,\Sb_i)$ for $i=1,\ldots,K$.
Moreover, assume that each receiver $i$ decodes $\xb_i$ by single user detection in the ID mode. Then the achievable information rate of receiver $i$ is given by
\begin{align}\label{rate function}
  R_i(\Sb_1,\ldots,\Sb_K)
     &=\log_2\left(1+\frac{\hb_{ii}^H\Sb_i\hb_{ii}}{\sum_{k\neq i}\hb_{ki}^H\Sb_k\hb_{ki} + \sigma_i^2}\right),
\end{align}
for $i=1,\ldots,K$.
Alternatively, the receiver may choose to harvest energy from the received signal. It can be
assumed that the total harvested RF-band energy during a transmission interval $\Delta$ is proportional to the power of the received baseband signal \cite{Zhangrui2011}. Specifically, for receiver $i$, the harvested energy, denoted by $\mathcal{E}_i$, can be expressed as
\begin{align}\label{energy function}
  \mathcal{E}_i= \gamma \Delta \sum_{k=1}^K&\hb_{ki}^H\Sb_k\hb_{ki},~i=1,\ldots,K,
\end{align}
where $\gamma$ is a constant accounting for the energy conversion loss in the transducer \cite{Zhangrui2011}.

Suppose that the receivers desire to harvest certain amounts of energy.
We are interested in investigating the optimal transmission strategies of $\Sb_i,$ $i=1,\ldots,K$, so that the information throughput of the $K$-user IFCs can be maximized while the energy harvesting requirements of the receivers are satisfied at the same time.
One should note that current energy harvesting receivers are not yet able to decode the information bits simultaneously \cite{Zhangrui2011}.
In subsequent sections, we will first study an ``ideal" scenario where the receivers can simultaneously operate in the ID mode and EH mode. Then, we further investigate some practical schemes where the receivers operate either in the ID mode or EH mode at any time instant.
In order to gain more insights, we will begin our investigation with the two-user scenario ($K=2$), and later extend the studies to the general $K$-user case (in Section \ref{sec: k user}).

\vspace{-3mm}
\section{Optimal WIET Design for Ideal Scheme}\label{sec:ideal receiver}

Let us assume that $K=2$ and consider ideal receivers which can simultaneously decode the information bits and harvest the energy from the received signals.
Suppose that the two receivers desire to harvest total amounts of energy $E_1$ and $E_2$, respectively.
We are interested in the following transmitter design problem for WIET:\vspace{-1mm}
\begin{subequations}\label{(P)}
    \begin{align}
        {\sf (P)}~~\max_{\Sb_1\succeq\zerob,\,\Sb_2\succeq\zerob}~
            &w_1 R_1(\Sb_1,\Sb_2)+w_2 R_2(\Sb_1,\Sb_2)\label{(P)_a}\\
        \st~&\hb_{11}^H\Sb_1\hb_{11} + \hb_{21}^H\Sb_2\hb_{21} \geq E_1,\label{(P)_b}\\
            &\hb_{22}^H\Sb_2\hb_{22} + \hb_{12}^H\Sb_1\hb_{12} \geq E_2,\label{(P)_c}\\
            &\Tr(\Sb_1) \leq P_1,\label{(P)_d}\\
            &\Tr(\Sb_2) \leq P_2,\label{(P)_e}
    \end{align}
\end{subequations}
where $w_1,w_2>0$ are positive weights, and $P_1>0$ and $P_2>0$ in \eqref{(P)_d} and \eqref{(P)_e} represent the individual power constraints. The constraints in \eqref{(P)_b} and \eqref{(P)_c} are the energy harvesting constraints where we have set $\gamma=\Delta=1$ for notational simplicity. Note that, in the absence of \eqref{(P)_b} and \eqref{(P)_c}, problem {\sf (P)} reduces to the classical sum rate maximization problem in MISO IFC \cite{Jorswieck08}:
\begin{subequations}\label{SRM IFC}
\begin{align}
\max_{\Sb_1\succeq\zerob,\,\Sb_2\succeq\zerob}~
    &w_1 R_1(\Sb_1,\Sb_2)+w_2 R_2(\Sb_1,\Sb_2)\label{SRM IFC_a}\\
\st~&\Tr(\Sb_1) \leq P_1,\label{SRM IFC_d}\\
    &\Tr(\Sb_2) \leq P_2.\label{SRM IFC_e}
\end{align}
\end{subequations}

It can be observed from \eqref{(P)} and \eqref{SRM IFC} that the energy harvesting constraints \eqref{(P)_b} and \eqref{(P)_c} would trade the maximum achievable sum rate for energy harvesting; i.e., the maximum sum rate in \eqref{(P)_a} is in general no larger than that in \eqref{SRM IFC_a}. To see when this would happen, let $(\bar\Sb_1^\star,\bar\Sb_2^\star)$ be an optimal solution to problem \eqref{SRM IFC}. One can verify from the rate function in \eqref{rate function} and problem \eqref{SRM IFC} that $(\bar\Sb_1^\star,\bar\Sb_2^\star)$ must satisfy
\ifconfver
\begin{align}\label{power region}
  &\!\!\!\!\begin{bmatrix}\hb_{11}^H\bar\Sb_1^\star\hb_{11}\\\hb_{12}^H\bar\Sb_1^\star\hb_{12}\end{bmatrix}
           \in\Omega_1\triangleq\bigg\{\begin{bmatrix}E_{11}&E_{12}\end{bmatrix}^T \bigg|\notag\\
  &\,E_{11}\!=\!\!\max_{\substack{\Sb_1\succeq\zerob,\Tr(\Sb_1)\leq P_1,\\\hb_{12}^H\Sb_1\hb_{12}\leq E_{12}}}
   \!\!\hb_{11}^H\Sb_1\hb_{11},\,0\leq E_{12}\!\leq\!P_1\|\hb_{12}\|^2\bigg\},\!
\end{align}
\begin{align}
  &\!\!\!\!\begin{bmatrix}\hb_{21}^H\bar\Sb_2^\star\hb_{21}\\\hb_{22}^H\bar\Sb_2^\star\hb_{22}\end{bmatrix}
           \in\Omega_2\triangleq\bigg\{\begin{bmatrix}E_{21}&E_{22}\end{bmatrix}^T \bigg|\notag\\
  &\,E_{22}\!=\!\!\max_{\substack{\Sb_2\succeq\zerob,\Tr(\Sb_2)\leq P_2,\\\hb_{21}^H\Sb_2\hb_{21}\leq E_{21}}}
   \!\!\hb_{22}^H\Sb_2\hb_{22},\,0\leq E_{21}\!\leq\!P_2\|\hb_{21}\|^2\bigg\}.\!\label{power region 2}
\end{align}
\else
\begin{align}\displaystyle\label{power region}
  &\begin{bmatrix}
     \hb_{11}^H\bar\Sb_1^\star\hb_{11}\\
     \hb_{12}^H\bar\Sb_1^\star\hb_{12}
  \end{bmatrix} \in \Omega_1 \triangleq \bigg\{
  \begin{bmatrix}E_{11}\\E_{12}\end{bmatrix}
  \bigg|~ E_{11}=\max_{\substack{\Sb_1\succeq\zerob,\Tr(\Sb_1) \leq P_1,
  \\\hb_{12}^H\Sb_1\hb_{12}\leq E_{12}}} \hb_{11}^H\Sb_1\hb_{11},~ 0\leq E_{12}\leq P_1\|\hb_{12}\|^2
  \bigg\},
  \\
  &\begin{bmatrix}
     \hb_{21}^H\bar\Sb_2^\star\hb_{21}\\
     \hb_{22}^H\bar\Sb_2^\star\hb_{22}
  \end{bmatrix} \in \Omega_2 \triangleq \bigg\{
  \begin{bmatrix}E_{21}\\E_{22}\end{bmatrix}
  \bigg|~ E_{22}=\max_{\substack{\Sb_2\succeq\zerob,\Tr(\Sb_2) \leq P_2,
  \\\hb_{21}^H\Sb_2\hb_{21}\leq E_{21}}} \hb_{22}^H\Sb_2\hb_{22},~ 0\leq E_{21}\leq P_2\|\hb_{21}\|^2
  \bigg\}. \label{power region 2}
\end{align}
\fi
That is, the energies harvested at the two receivers due to $(\bar\Sb_1^\star,\bar\Sb_2^\star)$ must lie in $\Omega_1+\Omega_2$. It can be shown that in $\Omega_1+\Omega_2$,
\ifconfver
\begin{subequations}\label{min energy}
\begin{align}
    \hb_{11}^H\bar\Sb_1^\star\hb_{11}\!+\!\hb_{21}^H\bar\Sb_2^\star\hb_{21}
   &\geq\min_{(E_{11},E_{12})\in\Omega_1,(E_{21},E_{22})\in \Omega_2}\!\!E_{11}+E_{21}\notag\\
   &= P_1 \|\hb_{11}^H\hat\hb_{12}^\perp\|^2,\label{min energy 1}\\
   \hb_{22}^H\bar\Sb_2^\star\hb_{22}\!+\!\hb_{12}^H\bar\Sb_1^\star\hb_{12}
   &\geq\min_{(E_{11},E_{12})\in \Omega_1,(E_{21},E_{22})\in \Omega_2}\!\!E_{22}+E_{12}\notag\\
   &= P_2 \|\hb_{22}^H\hat\hb_{21}^\perp\|^2,   \label{min energy 2}
\end{align}
\end{subequations}
\else
\begin{subequations}\label{min energy}
\begin{align}
   \hb_{11}^H\bar\Sb_1^\star\hb_{11}+\hb_{21}^H\bar\Sb_2^\star\hb_{21}&\geq \min_{(E_{11},E_{12})\in \Omega_1,~(E_{21},E_{22})\in \Omega_2}~E_{11}+E_{21} =P_1 \|\hb_{11}^H\hat\hb_{12}^\perp\|^2, \label{min energy 1}\\
    \hb_{22}^H\bar\Sb_2^\star\hb_{22}+\hb_{12}^H\bar\Sb_1^\star\hb_{12}&\geq \min_{(E_{11},E_{12})\in \Omega_1,~(E_{21},E_{22})\in \Omega_2}~E_{22}+E_{12} =P_2 \|\hb_{22}^H\hat\hb_{21}^\perp\|^2,   \label{min energy 2}
\end{align}
\end{subequations}
\fi
where $\hat\hb_{ij}^\perp\triangleq \frac{\Pi_{\hb_{ij}}^\perp\hb_{ii}}{\|\Pi_{\hb_{ij}}^\perp\hb_{ii}\|}$. Equations in \eqref{min energy} implies that the two receivers can at lease harvest energies $P_1 \|\hb_{11}^H\hat\hb_{12}^\perp\|^2$ and $P_2 \|\hb_{22}^H\hat\hb_{21}^\perp\|^2$, respectively.
The minimum amounts of energies are achieved when $E_{11}=P_1 \|\hb_{11}^H\hat\hb_{12}^\perp\|^2$, $E_{12}=0$, $E_{22}=P_2 \|\hb_{22}^H\hat\hb_{21}^\perp\|^2$ and $E_{21}=0$; that is, when each of the transmitters only focus on transmitting signals to its own receiver, without allowing any leakage of energy to the other receiver. According to \eqref{min energy}, we have that
\begin{Property}\label{Property energy}
 The energy harvesting constraints \eqref{(P)_b} and \eqref{(P)_c} are inactive at the optimum if $E_1\leq P_1 \|\hb_{11}^H\hat\hb_{12}^\perp\|^2$ and $E_2\leq P_2 \|\hb_{22}^H\hat\hb_{21}^\perp\|^2$; hence,
 {\sf (P)} reduces to the conventional MISO IFC problem \eqref{SRM IFC} under this condition.
\end{Property}

However, when $E_1> P_1 \|\hb_{11}^H\hat\hb_{12}^\perp\|^2$ or $E_2>  P_2 \|\hb_{22}^H\hat\hb_{21}^\perp\|^2$, the maximum information throughput may have to be compromised with energy harvesting.
Interestingly, the following proposition shows that the optimal transmit structure of {\sf(P)} is still similar to problem \eqref{SRM IFC} which does not have the energy harvesting constraints.

\begin{Proposition}\label{proposition:(P)_rnk1}
Assume that problem {\sf (P)} is feasible, and that $\hb_{11}\nparallel\hb_{12}$ and $\hb_{21}\nparallel\hb_{22}$ without loss of generality. Let $(\Sb_1^\star,\Sb_2^\star)$ denote the optimal solution to problem {\sf (P)}. Then, $\Tr(\Sb_1^\star)=P_1$ and $\Tr(\Sb_2^\star)=P_2$. Moreover, there exist { $a_i\in\Rset,~b_i \in \mathbb{C}$}, $i=1,2$, such that
\begin{subequations}\label{optimal S}
\begin{align}
   \Sb_1^\star= (a_1 \hb_{11} + b_1 \hb_{12})(a_1 \hb_{11} + b_1 \hb_{12})^H,\\
   \Sb_2^\star= (a_2 \hb_{21} + b_2 \hb_{22})(a_2 \hb_{21} + b_2 \hb_{22})^H.
\end{align}
\end{subequations}
\end{Proposition}

The proof is given in Appendix \ref{appendix: proof of prop 1}. Proposition \ref{proposition:(P)_rnk1} implies that beamforming is an optimal transmission strategy of {\sf(P)}. Moreover, the beamforming direction of transmitter $i$ should lie in the range space of $[\hb_{i1}, \hb_{i2}]$, for $i=1,2$, which is the same as the optimal beamforming direction of problem \eqref{SRM IFC} in the conventional IFCs \cite{Jorswieck08}. Given \eqref{optimal S}, the search of $\Sb_1$ and $\Sb_2$ in {\sf (P)} reduces to the search of $a_i$ and $b_i$ over the ellipsoids $\|a_i\hb_{i1}+b_i\hb_{i2}\|^2=P_i$ for all $i=1,2.$ However, unlike problem \eqref{SRM IFC}, optimizing the coefficients $a_i,b_i$, $i=1,2$, for problem {\sf(P)} have to take into account both the needs of energy harvesting and information transfer.

\begin{Remark}
    {\rm It is important to remark that, while {\sf (P)} is ideal in the sense that the receivers can simultaneously operate in the ID and EH modes, {\sf (P)} does not necessarily perform best in terms of sum rate maximization. The reason is that the cross-link signal power $\hb_{ik}^H\Sb_i\hb_{ik}$ plays two completely opposite roles in the considered scenario -- It can boost the energy harvesting of receiver $k$ on one hand, but also degrades the achievable information rate on the other hand. Therefore, when the cross-link channel power is strong (e.g., the interference dominated scenario) and when the energy harvesting constraints are not negligible (e.g., the conditions in Property \ref{Property energy} do not hold), the transmitters have to compromise the achievable information rate for energy harvesting. Under such circumstances, it might be a wiser strategy to split the ID and EH modes in time.}
\end{Remark}

\ifconfver
\begin{figure}
 \centering
 \subfloat[Sum rate vs. EH requirement $E$, for $N_t=4$ and $\SNR=10$ dB. Parameter $\eta$ measures the cross-link channel power.]
          {\label{fig1:a}\includegraphics[width=0.75\linewidth]{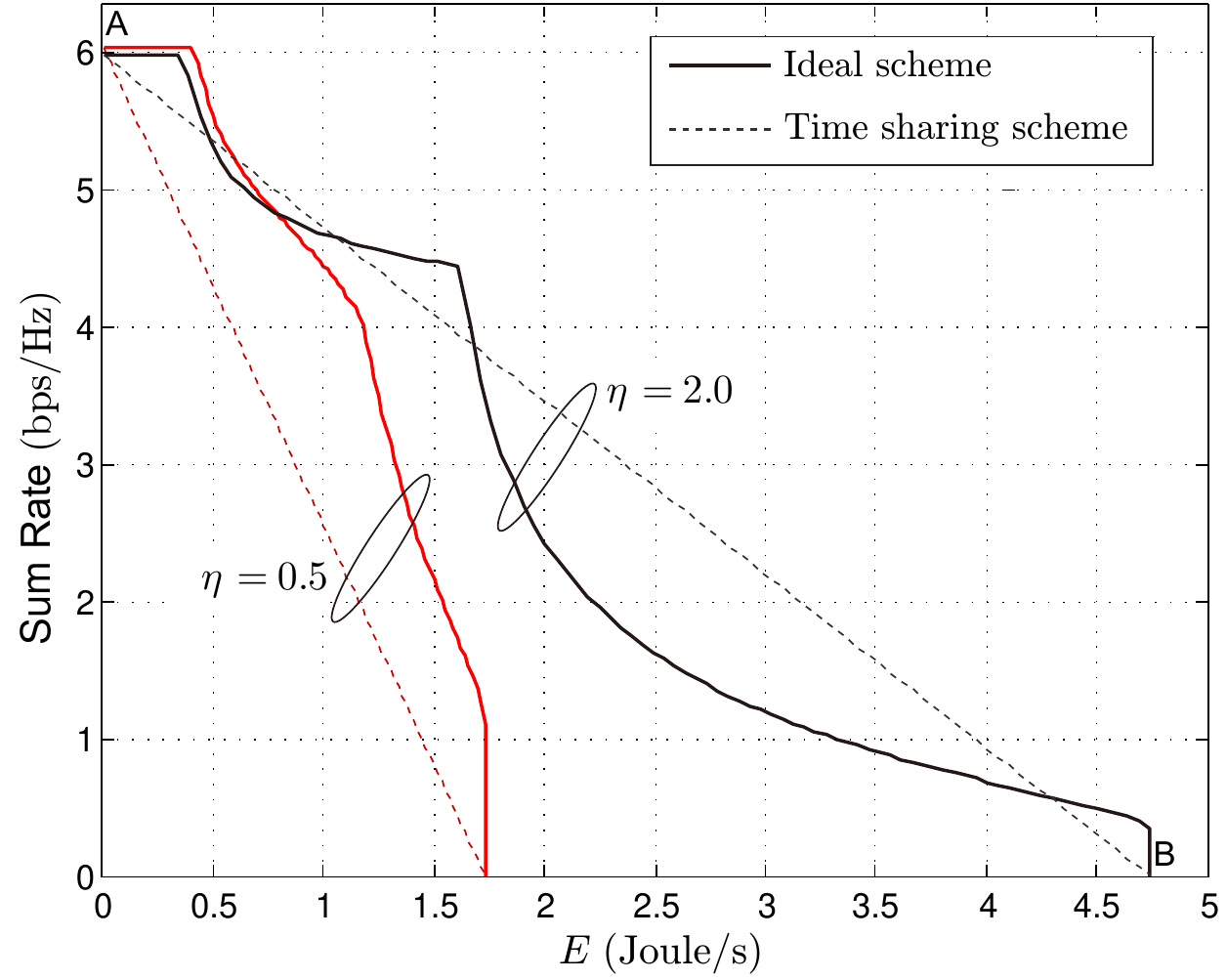}}\\
 \subfloat[Achievable rate region $(R_1,R_2)$, for $N_t=4$, $E_1=3$, $E_2=1$, $\eta=2$ and $\SNR=10$ dB.]
          {\label{fig1:b}\includegraphics[width=0.75\linewidth]{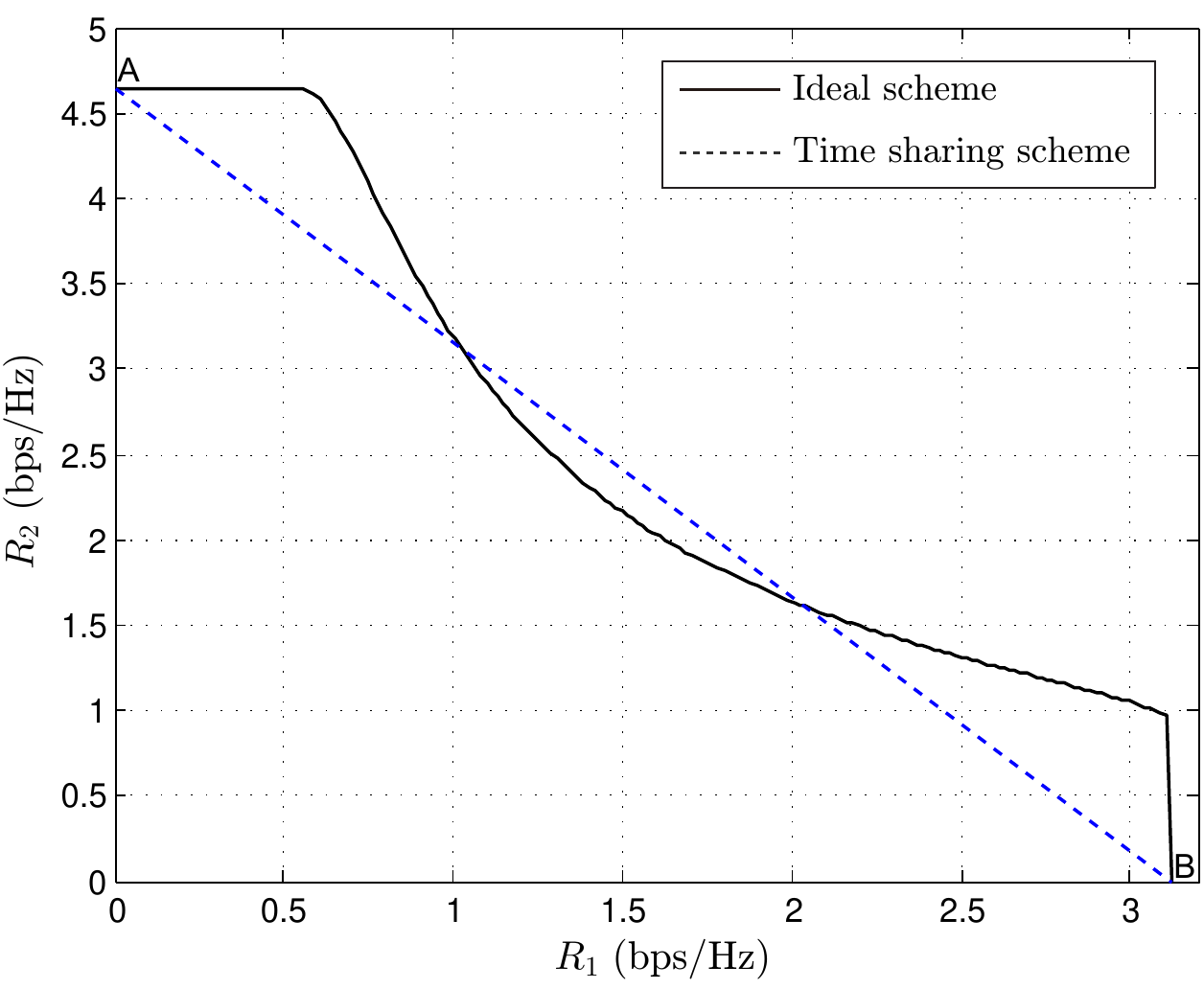}}
 \caption{Motivating simulation examples for the 2-user scenario.} 
 \label{fig1}\vspace{-5mm}
\end{figure}
\else
\begin{figure}
 \centering
 \subfloat[Sum rate vs. EH requirement $E$, for $N_t=4$ and $\SNR=10$ dB. Parameter $\eta$ measures the cross-link channel power.]
          {\label{fig1:a}\includegraphics[width=0.49\linewidth]{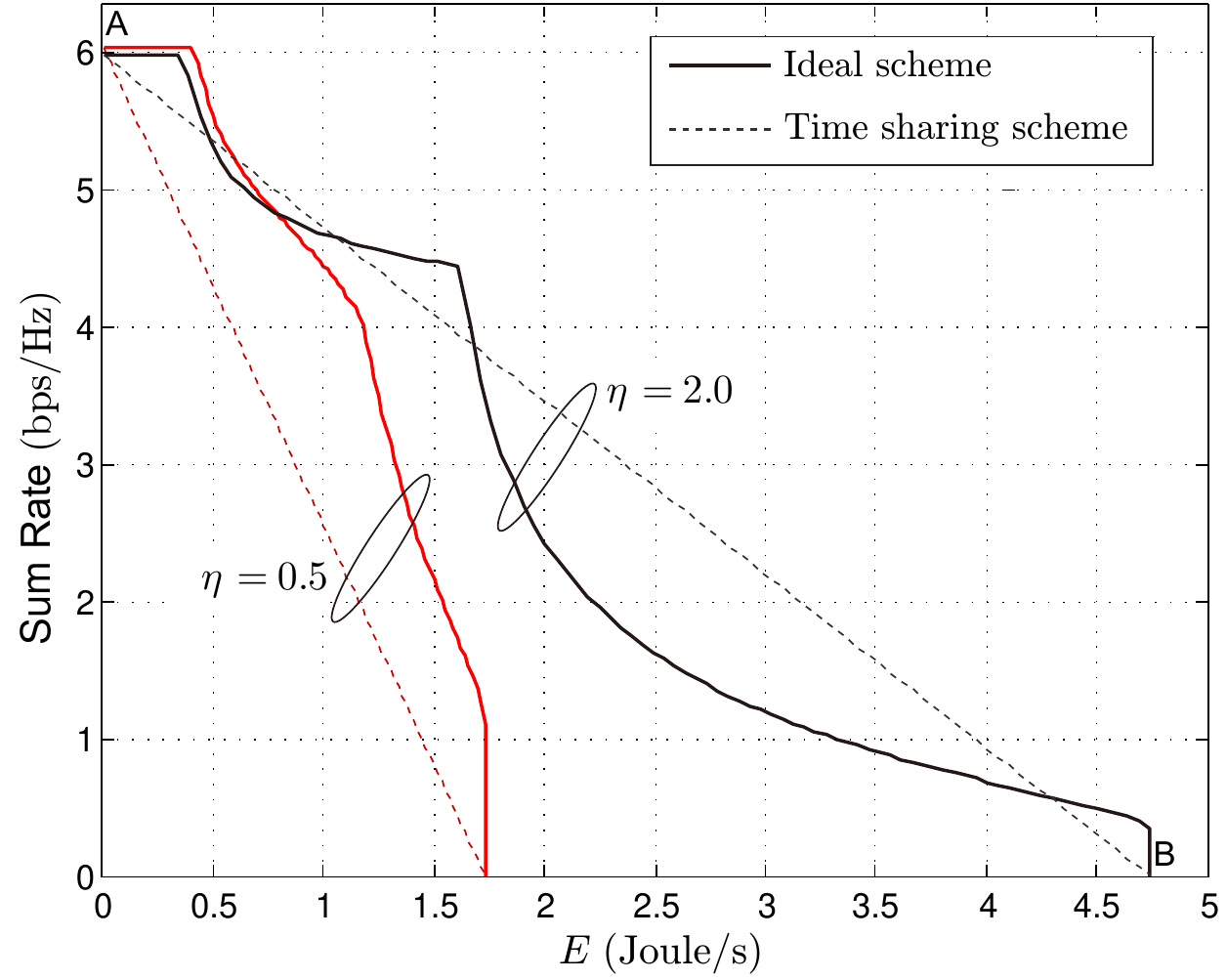}}~
 \subfloat[Achievable rate region $(R_1,R_2)$, for $N_t=4$, $E_1=3$, $E_2=1$, $\eta=2$ and $\SNR=10$ dB.]
          {\label{fig1:b}\includegraphics[width=0.49\linewidth]{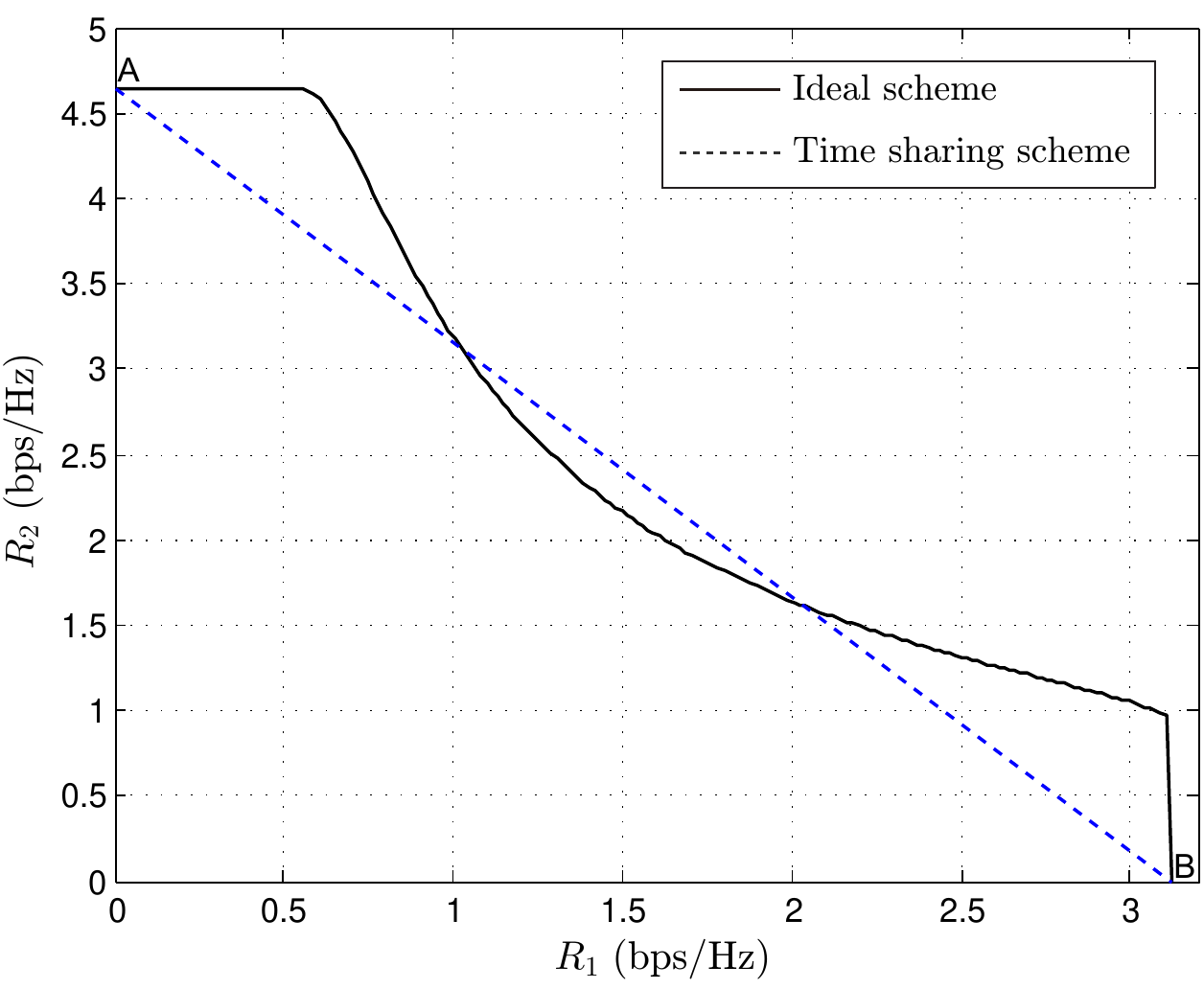}}
 \caption{Motivating simulation examples for the 2-user scenario ($K=2$).}
 \label{fig1}\vspace{-0.8cm}
\end{figure}
\fi
To further look into this aspect, we present in Fig. \ref{fig1} two simulation examples for the 2-user scenario. The detailed setting of the simulations are presented in Section \ref{sec: simulation}.
Fig. \ref{fig1:a} shows the sum rate-versus-energy requirement regions for two randomly generated channel realizations. The curves are obtained by exhaustively solving {\sf (P)} for various values of symmetric energy requirement $E\!\triangleq \!E_1 \!=\!E_2$.
The average powers of the direct link channels are normalized to one, while the average powers of the cross-link channels are measured by the parameter $\eta$. As one can observe from this figure, for $\eta=2$, the rate-energy region is not convex for this randomly generated channel realization. Moreover, for some values of $E$, the receivers may achieve a higher sum rate through time sharing between the EH mode and ID mode (see the dashed line between point A and point B). Fig. \ref{fig1:b} displays the rate region ($R_1$ versus $R_2$) of the two users. Analogously, we observe that time sharing for multiple access may achieve a higher sum rate (see the dashed line between points A and B).

The two simulation results in Fig. \ref{fig1} imply that the ideal scheme {(\sf P)} may not always achieve the best tradeoff between information transfer and energy harvesting, but, instead, time sharing for EH/ID mode switching or time sharing for multiple access may yield higher information sum rate. This motivates us to develop two practical schemes, namely, the {\it time-division mode switching (TDMS)} scheme and the {\it time-division multiple access (TDMA)} scheme, in the next section.
It is worthwhile to note that, in these time sharing schemes, the receivers operate either in the EH mode or ID mode at each time instant, and thus are more practical than the ideal receivers.

\section{Practical WIET Schemes and Optimal Transmission Strategies}\label{sec:prac_trans}
\subsection[TDMS Scheme]{Time Division Mode Switching (TDMS) Scheme}\label{subsec:TDMA_A}
In the first practical scheme, we divide the transmission interval into two time slots. In one time slot, both receivers operate in the EH mode, whereas, in the other time slot, both receivers switch to the ID mode. The two receivers thus coherently switch between the EH and ID modes, i.e., mode switching. Suppose that $\alpha$ fraction of the time is for EH mode and $(1-\alpha)$ fraction of the time is for ID mode. The TDMS scheme is described as follows:
\begin{itemize}
\item Time slot 1 (EH mode): The two receivers focus on harvesting the required energy $E_1$ and $E_2$ in $\alpha$ fraction of the time, i.e.,
    \begin{subequations}\label{eq: energy requirement}
    \begin{align}
    &\alpha\cdot(\hb_{11}^H\Sb_1\hb_{11}+\hb_{21}^H\Sb_2\hb_{21})\ge{E_1},\\
        &\alpha\cdot(\hb_{22}^H\Sb_2\hb_{22}+\hb_{12}^H\Sb_1\hb_{12})\ge{E_2}.
    \end{align}
    \end{subequations}
\item Time slot 2 (ID mode): Both the two receivers operate in the ID mode and maximize the information throughput in the remaining fraction of the time, i.e.,
    \begin{subequations}\label{TDMA_A_ID}
    \begin{align}\!\!\!\!\!\!\!\!\!\!
    \max_{\Sb_1\succeq\zerob,\,\Sb_2\succeq\zerob}~&\!\!(1-\alpha)\left(w_1 R_1(\Sb_1,\Sb_2)\!+\!
    w_2R_2(\Sb_1,\Sb_2)\right) \!\!\!\!\!\\
    \st~~&\Tr(\Sb_1)\leq P_1,~\Tr(\Sb_2)\leq P_2.
    \end{align}
    \end{subequations}
\end{itemize}
Problem \eqref{TDMA_A_ID} in the ID mode is the classical sum rate maximization problem in the MISO IFC [see \eqref{SRM IFC}], which can be efficiently handled by existing methods in \cite{Jorswieck08,Lindblom2011,Liu2012_IFC}. Note that it has been shown in \cite{ZhangCui2010,Shang2011} that beamforming is an optimal transmission scheme for problem \eqref{TDMA_A_ID}.

We now focus on the EH mode in time slot 1. Since time slot 1 does not contribute to the information throughput, it is desirable to spend as
least as possible time for the EH mode, i.e., to use a minimal time fraction $\alpha$ to fulfill the energy harvesting task. Mathematically, we can write it as the following optimization problem
\begin{subequations}\label{TDMA_A_EH}
\begin{align}
    \max_{\beta\in \mathbb{R},\,\Sb_1\succeq\zerob,\,\Sb_2\succeq\zerob}&~~\beta\label{TDMA_A_EH_a}\\
    \st~~~&\!\!\hb_{11}^H\Sb_1\hb_{11}+\hb_{21}^H\Sb_2\hb_{21}\ge\beta{E_1},\label{TDMA_A_EH_b}\\
    &\!\!\hb_{12}^H\Sb_1\hb_{12}+\hb_{22}^H\Sb_2\hb_{22}\ge\beta{E_2},\label{TDMA_A_EH_c}\\
    &\!\!\!\Tr(\Sb_1)\le{P_1},~\Tr(\Sb_2)\le{P_2},\label{TDMA_A_EH_d}
\end{align}
\end{subequations}
where $\beta\triangleq1/\alpha$. Note that if the optimal $\beta$ of \eqref{TDMA_A_EH} is less than one (i.e., optimal $\alpha> 1$), then it implies that the energy harvesting requirements \eqref{eq: energy requirement} cannot be satisfied even if the receivers dedicate themselves to harvesting energy throughout the whole transmission interval. In that case, we declare that the TDMS scheme is not feasible.

While problem \eqref{TDMA_A_EH} is a convex semidefinite program (SDP), which can be solved by the off-the-shelf solvers, we show that \eqref{TDMA_A_EH} actually admits a semi-analytical solution:

\begin{Proposition}\label{proposition: one dimensional prob}
Assume that $\hb_{i1}$ and $\hb_{i2}$ are linearly independent but not orthogonal to each other, for $i=1,2$. The optimal solution to problem \eqref{TDMA_A_EH} is given by
\ifconfver
\begin{subequations}\label{solution of TDMA A EH}
\begin{align}
   \Sb_1(\mu^\star)&=P_1\vb_1(\mu^\star)\vb_1^H(\mu^\star),~\Sb_2(\mu^\star)
                    =P_2\vb_2(\mu^\star)\vb_2^H(\mu^\star),\label{solution of TDMA A EH 1}\\
   \beta(\mu^\star)&=\min
    \left\{\frac{\hb_{11}^H\Sb_1(\mu^\star)\hb_{11}+\hb_{21}^H\Sb_2(\mu^\star)\hb_{21}}{E_1},\right.\notag\\
   &~~~~~~~~~~~~~
    \left. \frac{\hb_{12}^H\Sb_1(\mu^\star)\hb_{12}\!+\!\hb_{22}^H\Sb_2(\mu^\star)\hb_{22}}{E_2}\right\},\!\!\!\label{solution of TDMA A EH 2}
\end{align}
\end{subequations}
\else
\begin{subequations}\label{solution of TDMA A EH}
\begin{align}
&\Sb_1(\mu^\star)=P_1\vb_1(\mu^\star)\vb_1^H(\mu^\star),~\Sb_2(\mu^\star)
=P_2\vb_2(\mu^\star)\vb_2^H(\mu^\star),\label{solution of TDMA A EH 1}\\
&\beta(\mu^\star)=\min\left
\{\frac{\hb_{11}^H\Sb_1(\mu^\star)\hb_{11}+\hb_{21}^H\Sb_2(\mu^\star)
\hb_{21}}{E_1},~\frac{\hb_{12}^H\Sb_1(\mu^\star)\hb_{12}+\hb_{22}^H\Sb_2(\mu^\star)
\hb_{22}}{E_2}\right\},\label{solution of TDMA A EH 2}
\end{align}
\end{subequations}
\fi
where $\mu^\star\geq 0$ is the optimal dual variable associated with constraint \eqref{TDMA_A_EH_b}, and $\vb_i(\mu^\star)$ is the principal eigenvector of $\mu^\star\hb_{i1}\hb_{i1}^H+\frac{(1-\mu^\star E_1)}{E_2}\hb_{i2}\hb_{i2}^H$ for $i=1,2$.
Moreover, $\mu^\star$ can be efficiently obtained using a simple bisection search.
\end{Proposition}

The proof of Proposition \ref{proposition: one dimensional prob} is given in Appendix \ref{appendix: proof of prop 2}.
The assumptions on $\hb_{i1}$ and $\hb_{i2}$, for $i=1,2$, hold with probability one for random (continuous) fading channels.
Note that Proposition \ref{proposition: one dimensional prob} also implies that beamforming is optimal for the EH mode of the TDMS scheme.

\vspace{-2mm}
\subsection{TDMA Scheme}\label{subsec:TDMA_B}
Unlike TDMS scheme, in each time slot of TDMA scheme, one receiver operates in the ID mode and the other receiver operates in the EH mode. Assume that the time fraction of the first time slot is $\alpha$.
\begin{itemize}
\item Time slot 1: Receiver 1 operates in the ID mode and receiver 2 operates in the EH mode. The objective is to maximize the information rate of receiver 1 and guarantee the energy harvesting requirement of receiver 2 at the same time. The design problem is given by
\begin{subequations}\label{TDMA_B_1}
\begin{align}
\max_{\Sb_1\succeq\zerob,\,\Sb_2\succeq\zerob}~&\alpha\log_2\left(1+\frac{\hb_{11}^H\Sb_1\hb_{11}}{\hb_{21}^H\Sb_2\hb_{21}+\sigma_1^2}\right)\label{TDMA_B_1_a}\\
\st~&\hb_{12}^H\Sb_1\hb_{12}+\hb_{22}^H\Sb_2\hb_{22}\ge{E_2}/\alpha,\label{TDMA_B_1_b}\\
&\Tr(\Sb_1)\le{P_1},~\Tr(\Sb_2)\le{P_2},\label{TDMA_B_1_c}
\end{align}
\end{subequations}
\item Time slot 2: The operation modes of the two receivers are exchanged:
\begin{subequations}\label{TDMA_B_2}
\begin{align}
\max_{\Sb_1\succeq\zerob,\,\Sb_2\succeq\zerob}~&(1-\alpha)\log_2\left(1+\frac{\hb_{22}^H\Sb_2\hb_{22}}{\hb_{12}^H\Sb_1\hb_{12}+\sigma_2^2}\right)\\
\st~~~~&\!\!\!\!\hb_{11}^H\Sb_1\hb_{11}+\hb_{21}^H\Sb_2\hb_{21}\!\ge\!{E_1}/(1-\alpha),\label{TDMA_B_2_b}\\
&\!\!\!\Tr(\Sb_1)\le{P_1},~\Tr(\Sb_2)\le{P_2}.
\end{align}
\end{subequations}
\end{itemize}
By intuition, this TDMA scheme would be of interest when the two receivers have asymmetric  energy harvesting requirements and asymmetric cross-link channel powers. Moreover, like the conventional interference channel without energy harvesting, the TDMA scheme may outperform the spectrum sharing schemes in interference dominated scenarios.
It is not difficult to show that:

\vspace{-2mm}
\begin{Lemma}\label{Lemma1}
The TDMA scheme is feasible if and only if
\begin{align}\label{TDMA-B_E-limit}
&\frac{E_1}{P_1\|\hb_{11}\|^2+P_2\|\hb_{21}\|^2}+\frac{E_2}{P_1\|\hb_{12}\|^2+P_2\|\hb_{22}\|^2}
\le 1.
\end{align}
\end{Lemma}

\emph{\bf Proof:} The TDMA scheme is feasible if and only if both \eqref{TDMA_B_1} and \eqref{TDMA_B_2} are feasible. Problem \eqref{TDMA_B_1} is feasible if and only if there exists some $\alpha\in[0,1]$ such that
\begin{align}
E_2&\le\alpha\cdot
\begin{pmatrix}
  \max\limits_{\Sb_1\succeq\zerob,\Sb_2\succeq\zerob}&\!\!\!\hb_{12}^H\Sb_1\hb_{12}+\hb_{22}^H\Sb_2\hb_{22}\\
  {^{\Tr(\Sb_1)\le{P_1},\Tr(\Sb_2)\le{P_2}}}&
\end{pmatrix} \notag \\
&=\alpha\cdot(P_1\|\hb_{12}\|^2+P_2\|\hb_{22}\|^2), \label{temp1}
\end{align} where the equality is obtained by applying the result in \cite[Proposition 2.1]{Zhangrui2011}.
Similarly, one can show that \eqref{TDMA_B_2} is feasible if and only if
\begin{align}
E_1\le(1-\alpha)\cdot(P_1\|\hb_{11}\|^2+P_2\|\hb_{21}\|^2).\label{temp2}
\end{align}
Combining \eqref{temp1} and \eqref{temp2} gives rise to \eqref{TDMA-B_E-limit}. Conversely, given \eqref{TDMA-B_E-limit}, let $\alpha\!=\!\frac{E_2}{P_1\|\hb_{12}\|^2+P_2\|\hb_{22}\|^2}$, and thus
$\frac{E_1}{P_1\|\hb_{11}\|^2+P_2\|\hb_{21}\|^2}\!\leq\! 1-\alpha$, which are \eqref{temp1} and \eqref{temp2}, respectively. Hence, when \eqref{TDMA-B_E-limit} is true, the TDMA scheme is feasible. \hfill $\blacksquare$

According to \eqref{temp1} and \eqref{temp2}, a feasible time fraction $\alpha$ must lie in the interval
\begin{align}\label{feasible alpha}\!
 \frac{E_2}{P_1\|\hb_{12}\|^2\!+\!P_2\|\hb_{22}\|^2}\!\leq\!\alpha\!\leq\!1\!-\! \frac{E_1}{P_1\|\hb_{11}\|^2\!+\!P_2\|\hb_{21}\|^2}.\!
\end{align}
Interestingly, given a feasible $\alpha$, both problems \eqref{TDMA_B_1} and \eqref{TDMA_B_2} can be efficiently solved (semi-analytically).
Since problems \eqref{TDMA_B_1} and \eqref{TDMA_B_2} are similar to each other, we take \eqref{TDMA_B_1} as the example.

\vspace{-2mm}
\begin{Proposition}\label{proposition: TDMA B}
Let the time fraction $\alpha$ satisfy \eqref{feasible alpha}. Then, an optimal solution to problem \eqref{TDMA_B_1}, denoted by ($\Sb_1^\star$, $\Sb_2^\star$), is given by
\begin{align}\label{X1X2}
  &\Sb_1^\star=\vb_1(y^\star)\vb^H_1(y^\star)/y^\star,~\Sb_2^\star=\vb_2(y^\star)\vb^H_2(y^\star)/y^\star,
 \end{align}
where\vspace{-2mm}
\ifconfver
\begin{align*}
\vb_1(y) &=\left\{\!\!\!\!\!\!\!\!\!
\begin{array}{ll}
 &\sqrt{y P_1}\bar\hb_{11} ,~~~~~~{\rm if~} g(y)/y\geq {E_2/{\alpha}\!-\!P_1|\bar\hb_{11}^H\hb_{12}|^2},\\[4pt]
 &\frac{\sqrt{y{E_2}/{\alpha} - g(y)}}{|\hb_{12}^H\hb_{11}|}\bar\hb_{12}^H\hb_{11}\bar\hb_{12} \\
 &~~~~~~~~~~~~+ \sqrt{yP_1\!-\!\frac{y{E_2}/{\alpha} - g(y)}{\|\hb_{12}\|^2}}\bar\hb_{12}^\perp,~  {\rm otherwise,}
\end{array}\right.\\[1pt]
\vb_2(y) &= \frac{\sqrt{1- y\sigma_1^2}}{|\hb_{21}^H\hb_{22}|}(\bar\hb_{21}^H\hb_{22})\bar\hb_{21}
           + \sqrt{y P_2-\frac{1- y\sigma_1^2}{\|\hb_{21}\|^2}}\bar\hb_{21}^\perp,
\end{align*}\vspace{-8pt}
\begin{gather}\!\!\!\!\!\!\!\!\!\!\!\!\!\!\!\!\!\!\!\!\!\!\!\!\!\!
\begin{split}
  y^\star&= \argmax_{y}~|\hb_{11}^H\vb_1(y)|^2\\[-4pt]
         &~~~~~~~~~\st~\frac{1}{P_2|\hb_{21}^H\bar\hb_{22}|^2+\sigma_1^2} \leq y \leq \frac{1}{\sigma_1^2},\label{Eq:optimal y00}
\end{split}
\end{gather}
\else
\begin{align}
\vb_1(y) &=\left\{
\begin{array}{ll}
 &\sqrt{y P_1}\bar\hb_{11} ,~~~~~~~~~~~~~~~~~~~~~~~~~~{\rm if~} g(y)/y\geq {E_2/{\alpha}-P_1|\bar\hb_{11}^H\hb_{12}|^2},\\
 &\frac{\sqrt{y{E_2}/{\alpha} - g(y)}}{|\hb_{12}^H\hb_{11}|}\bar\hb_{12}^H\hb_{11}\bar\hb_{12}+ \sqrt{yP_1\!-\!\frac{y{E_2}/{\alpha} - g(y)}{\|\hb_{12}\|^2}}\bar\hb_{12}^\perp,~  {\rm otherwise,}
\end{array}
\right. \label{v1 0}\\
\vb_2(y) &= \frac{\sqrt{1- y\sigma_1^2}}{|\hb_{21}^H\hb_{22}|}(\bar\hb_{21}^H\hb_{22})\bar\hb_{21}
           + \sqrt{y P_2-\frac{1- y\sigma_1^2}{\|\hb_{21}\|^2}}\bar\hb_{21}^\perp,\\
 \label{Eq:optimal y00}
  y^\star&= \arg ~\max_{y}~|\hb_{11}^H\vb_1(y)|^2 ~\st~\frac{1}{P_2|\hb_{21}^H\bar\hb_{22}|^2+\sigma_1^2} \leq y \leq \frac{1}{\sigma_1^2},
\end{align}
\fi
in which $\bar\hb_{ij}=\frac{\hb_{ij}}{\|\hb_{ij}\|}$, $\hat\hb_{ij}^{\bot}=\frac{\Pi_{\hb_{ij}}^\perp\bar\hb_{ii}} {\|\Pi_{\hb_{ij}}^\perp\bar\hb_{ii}\|}$ for $i=1,2,$ and $g(y) = |\hb_{22}^H\vb_2(y)|^2$. Problem \eqref{Eq:optimal y00} is a convex problem, and thus $y^\star$ can be obtained by a bisection search.
\end{Proposition}

The proof is presented in Appendix \ref{appendix of proof of prop 3}. We see from
\eqref{X1X2} that beamforming is also optimal to the TDMA scheme. By Proposition \ref{proposition: TDMA B}, given a feasible time fraction $\alpha$, one can efficiently solve problems \eqref{TDMA_B_1} and \eqref{TDMA_B_2} and thus evaluate the achievable sum rate of the two users. Then, the optimal time fraction $\alpha$ that maximizes the sum rate of the two users can be obtained by line search over the interval in \eqref{feasible alpha}.

\subsection{TDMA via Deterministic Signal for Energy Harvesting}\label{subsec:TDMA_C}
It should be noticed that, while Gaussian signaling is optimal for information transfer, it may not be necessary for energy transfer. In particular, if one user operates in the EH mode, the transmitter may simply transmit some deterministic signals (e.g., training/pilot signals) known to both receivers. Consider the TDMA scheme in the previous subsection, and assume that, in the first time slot, transmitter 2 operating in the EH mode transmits deterministic signals $\xb_2$ which are known to receiver 1 operating in the ID mode. Under such circumstances, receiver 1 can actually remove $\hb_{21}^H\xb_2$ from the received signal before information detection, i.e., removing the cross-link interference. The design problem in the 1st time slot thereby reduces to
\begin{subequations}\label{TDMA_C_1}
\begin{align}
\max_{\Sb_1\succeq\zerob,\,\Sb_2\succeq\zerob}~&\alpha\log_2\left(1+{\sigma_1^{-2}}\hb_{11}^H\Sb_1\hb_{11}\right)\label{TDMA_C_1_a}\\
\st~&\hb_{12}^H\Sb_1\hb_{12}+\hb_{22}^H\Sb_2\hb_{22}\ge{E_2}/\alpha,\label{TDMA_C_1_b}\\
&\Tr(\Sb_1)\le{P_1},~\Tr(\Sb_2)\le{P_2}.\label{TDMA_C_1_c}
\end{align}
\end{subequations}
Problem \eqref{TDMA_C_1} is easier to handle than its counterpart in \eqref{TDMA_B_1}.
Clearly, given $\alpha$ satisfying \eqref{feasible alpha}, optimal $\Sb_2$ is given by $\Sb_2^\star = P_2 \bar\hb_{22} \bar\hb_{22}^H,$
Therefore, \eqref{TDMA_C_1} boils down to
\begin{subequations}
    \begin{align}\!
        \max_{\Sb_1\succeq\zerob}~&\hb_{11}^H\Sb_1\hb_{11}\\
        \st~&\hb_{12}^H\Sb_1\hb_{12}\ge{E_2}/\alpha - P_2\|\hb_{22}\|^2,~\Tr(\Sb_1)\le{P_1},\!\!\!
    \end{align}
\end{subequations}
which admits a closed-form solution for $\Sb_1^\star$ according to \cite[Proposition 2.1]{Zhangrui2011}. Analogously, the design problem for the second time slot can be simplified. In this paper, we refer to this scheme as the \emph{TDMA (D) scheme}. Since the receivers are free from cross-link interference, it is anticipated that the TDMA (D) scheme performs no worse than the TDMA scheme. However, it should be noted that, in order to do so, the two receivers require perfect knowledge of the cross-link channels $\hb_{12}$ and $\hb_{21}$, respectively; otherwise, the receivers may suffer performance degradation due to imperfect interference cancelation.

We remark that, in addition to the above time sharing based schemes, it is also possible for the receivers to split the received signals into two parts, one for EH and the other for ID, i.e., power splitting (PS) \cite{Zhangrui2011}. This scheme will be studied in Section \ref{subsec: power splitting}.

\vspace{-2mm}
\section{WIET Design for $K$-user MISO IFC}\label{sec: k user}
In this section, we consider the WIET problem for the $K$-user MISO IFC scenario. We begin with the ideal scheme, and in the second subsection, we extend the TDMS and TDMA schemes in Section \ref{sec:prac_trans} to the $K$-user scenario. In the last subsection, we further investigate the PS scheme.

\vspace{-0.3cm}
\subsection{Transmitter Optimization for Ideal Receivers}\label{subsec: k user ideal}

By the signal model in \eqref{eq: received signal}, \eqref{rate function}, \eqref{energy function} and {\sf (P)} in \eqref{(P)}, the $K$-user WIET problem is formulated as
\begin{subequations}\label{Problem::Kuser::RateMax}
  \begin{align}
    \max_{\substack{\Sb_i\succeq\zerob \\ \forall i=1,\ldots,K} }~~&\sum_{i=1}^K w_i \log_2\left(1+\frac{\hb_{ii}^H\Sb_i\hb_{ii}}{\sum_{k\neq i}\hb_{ki}^H\Sb_k\hb_{ki} + \sigma_i^2}\right) \label{Problem::Kuser::RateMax:a}\\
    \st~~
    & \sum_{k=1}^K \hb_{ki}^H\Sb_k\hb_{ki}\geq E_i,~\forall i=1,\ldots,K,\label{Problem::Kuser::RateMax:c}\\
    & \Tr(\Sb_i) \leq P_i,~\forall i=1,\ldots,K,\label{Problem::Kuser::RateMax:d}
  \end{align}
\end{subequations}
where $E_i \geq 0$ is the energy requirement of user $i$, for $i=1,\ldots,K$. Since problem \eqref{Problem::Kuser::RateMax} is NP-hard in general \cite{Luo2008}, our interest for the $K$-user WIET problem lies in efficient approaches to finding an approximate solution.

We propose an efficient algorithm based on successive convex approximation (SCA) \cite{Marks1978} by adopting the log-exponential reformulation idea in \cite{WCLi2013}. Compared to the methods in \cite{Jorswieck08,Lindblom2011,Liu2012_IFC}, the proposed method can work for scenarios with a medium to large number of users. Specifically, by introducing slack variables $\{x_i,y_i\}$, we can reformulate problem \eqref{Problem::Kuser::RateMax} as\vspace{-1mm}
\begin{subequations}\label{Problem::Kuser::RateMax:slack}
  \begin{align}
    \max_{\substack{\Sb_i\succeq \zerob,\,x_i,\,y_i \\ \forall i=1,\ldots,K}}~~
    &\sum_{i=1}^K w_i{  (x_i - y_i)\log_2e} \label{Problem::Kuser::RateMax:slack:a}\\[-4pt]
    \st~~~~~~
    & \sum_{k=1    }^K\hb_{ki}^H\Sb_k\hb_{ki} + \sigma_i^2 \geq e^{x_i}~\forall i,\label{Problem::Kuser::RateMax:slack:b}\\[-3pt]
    & \sum_{k\neq i}^K\hb_{ki}^H\Sb_k\hb_{ki} + \sigma_i^2 \leq e^{y_i}~\forall i,\label{Problem::Kuser::RateMax:slack:c}\\[-3pt]
    & \eqref{Problem::Kuser::RateMax:c}, ~\eqref{Problem::Kuser::RateMax:d}.
  \end{align}
\end{subequations}
As seen, the rate functions in \eqref{Problem::Kuser::RateMax:a} are equivalently decomposed into the objective function in \eqref{Problem::Kuser::RateMax:slack:a} and the two constraints in \eqref{Problem::Kuser::RateMax:slack:b} and \eqref{Problem::Kuser::RateMax:slack:c}. In particular, one can verify that constraints \eqref{Problem::Kuser::RateMax:slack:b} and \eqref{Problem::Kuser::RateMax:slack:c} will hold with equality at the optimum, implying that \eqref{Problem::Kuser::RateMax:slack} is equivalent to \eqref{Problem::Kuser::RateMax}.

Problem \eqref{Problem::Kuser::RateMax:slack} has a linear objective function and convex constrains, except for constraint \eqref{Problem::Kuser::RateMax:slack:c}. We propose to linearly approximate constraint \eqref{Problem::Kuser::RateMax:slack:c} in an iterative manner. Suppose that, at iteration $n$, we are given $\Sb_1^\star[n-1],\ldots, \Sb_K^\star[n-1]$. Let $\bar y_i[n] \!=\! \ln\left(\sum_{k\neq i}^K \hb_{ki}^H\Sb_k^\star [n-1]\hb_{ki} + \sigma_i^2\right)$, $i=1,\ldots,K$. We solve the following problem at the $n$th iteration
\begin{subequations}\label{Problem::Kuser::RateMax:approx}
  \begin{align}\!\!\!
     &\!\!\!\!\!\!\!\!\!\!\!\!\!\!\!\!\!\!
     \{\Sb_i^\star[n]\}_{i=1}^K=
       \argmax_{\substack{\Sb_i\succeq \zerob,\,x_i,\,y_i\\ \forall i=1,\ldots,K}}~
       \sum_{i=1}^K w_i{  (x_i - y_i)}\log_2{e} \label{Problem::Kuser::RateMax:approx:a}\\[-2pt]
 \st~& \sum_{k=1}^K\hb_{ki}^H\Sb_k\hb_{ki} \!+\! \sigma_i^2 \geq e^{x_i}~\forall i,\label{approx:b}\\[-2pt]
     & \sum_{k\neq i}^K\hb_{ki}^H\Sb_k\hb_{ki}\!+\!\sigma_i^2\leq e^{\bar y_i[n] }(y_i\!-\!\bar y_i[n]\!+\!1)~\forall i,\label{approx:c}\\[-2pt]
     & \eqref{Problem::Kuser::RateMax:c},~\eqref{Problem::Kuser::RateMax:d}.
  \end{align}
\end{subequations}
Note that constraint \eqref{approx:c} is convex; it is a conservative approximation to \eqref{Problem::Kuser::RateMax:slack:c} since it holds that $e^{y_i}\geq e^{\bar y_i[n]}(y_i-\bar y_i[n] + 1)~\forall y_i$ due to the convexity of $e^{y_i}$.
As a result, problem \eqref{Problem::Kuser::RateMax:approx} is a convex SDP which can be solved efficiently by off-the-shelf solvers, e.g., \texttt{CVX} \cite{cvx}. Detailed steps of the proposed algorithm is summarized in Algorithm 1.

{\begin{algorithm}[h]\small
\caption{SCA algorithm for problem \eqref{Problem::Kuser::RateMax}}
\begin{algorithmic}[1]\label{alg:centralized}
\STATE {\bf Find} initial variables by solving the feasibility problem
  \begin{align*}
    \{\Sb_i^\star[0]\}_{i=1}^K ={\rm find}&~~\{\Sb_1,\ldots, \Sb_K\}\\
    \st &~\sum_{k=1}^K \hb_{ki}^H\Sb_k\hb_{iki}\geq E_i~\forall i,\\
        &~\Tr(\Sb_i) \leq P_i,~\Sb_i\succeq \zerob~\forall i.
  \end{align*}
If the problem is infeasible, then declare infeasibility of \eqref{Problem::Kuser::RateMax}; otherwise, set $n=0$ and perform the following steps.

\REPEAT
\STATE $n:=n+1$.
\STATE $\bar y_i[n] = \ln\left(\sum_{k\neq i}^K \hb_{ki}^H\Sb_k^\star[n-1]\hb_{ki} + \sigma_i^2\right)~\forall i.$
\STATE Solve problem \eqref{Problem::Kuser::RateMax:approx} to obtain $\{\Sb_1^\star[n],\ldots, \Sb_K^\star[n]\}$.
\UNTIL the stopping criterion is met.
\STATE {\bf Output}
  $(\Sb_1^\star[n],\ldots, \Sb_K^\star[n])$ as an approximate solution.
\end{algorithmic}
\end{algorithm}}

It can be shown that Algorithm 1 belongs to the category of the successive upper-bound minimization (SUM) method proposed in \cite{RazaviHongLuo2012} and can converge to a stationary point of problem \eqref{Problem::Kuser::RateMax}, as stated in Proposition \ref{proposition: conv of sca}. The details are relegated to Appendix \ref{Appendix proof of conv of sca}.

\begin{Proposition}\label{proposition: conv of sca}
  Any limit point of the sequence $\{\Sb_1^\star[n],\ldots, \Sb_K^\star[n]\}_{n=1}^\infty$ generated by Algorithm 1 is a stationary point of problem \eqref{Problem::Kuser::RateMax}.
\end{Proposition}

\vspace{-4mm}
\subsection{Practical $K$-User WIET Schemes }\label{subsec: k user TDMA}
We extend the TDMS and TDMA schemes in Section \ref{sec:prac_trans} to the general $K$-user scenario in this subsection.

\begin{figure}[!t]
 \centering
 \subfloat[TDMS]{\label{fig: time splitting a}\includegraphics[width=0.45\linewidth]{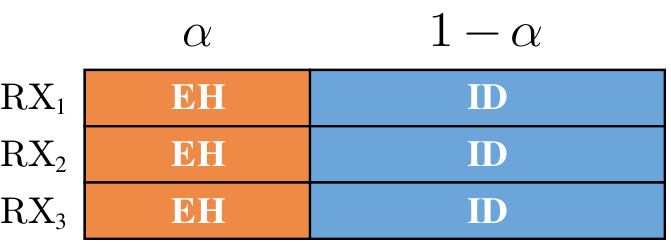}}~~
 \subfloat[TDMA]{\label{fig: time splitting b}\includegraphics[width=0.45\linewidth]{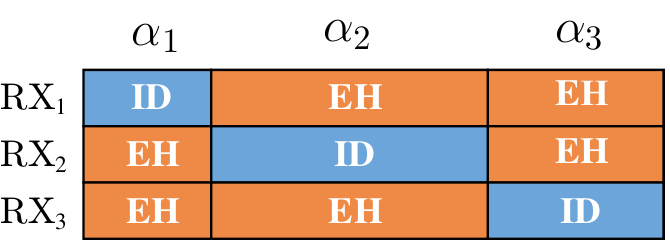}}
 \caption{Illustration of the proposed TDMS and TDMA schemes for WIET in a $3$-user scenario.}
 \label{fig: time splitting}\vspace{-3mm}
\end{figure}

{\bf 1) $K$-user TDMS scheme:} This scheme is similar to the TDMS scheme presented in Section \ref{subsec:TDMA_A}. In the 1st time slot, all users operate in the EH mode, and in the 2nd time slot, all users operate in the ID mode; see Fig. \ref{fig: time splitting a}. In the 1st time slot, the optimal time fraction $\alpha^\star$ and the associated optimal signal covariance matrices $\{\Sb_{k}^\star\}_{k=1}^K$
for energy harvesting can be obtained by solving a convex problem analogous to problem \eqref{TDMA_A_EH}. In the 2nd time slot, one has to solve the
classical sum rate maximization problem
\begin{subequations}\label{eq: SRM k user}
  \begin{align}\!\!\!\!\!
    \max_{\substack{\,\Sb_{i}\succeq\zerob,\\i=1,\ldots,K}}~
         &\!(1\!-\!\alpha)\sum_{i=1}^K\! w_i\log_2\!\!
         \left(\!1\!+\!\frac{\hb_{ii}^H\Sb_{i}\hb_{ii}}{\sum_{k\neq i}^K \hb_{ki}^H\Sb_{k}\hb_{ki}\!+\!\sigma_i^2}\right)\!\!\!\!\!\!\\
    \st~~ &\Tr(\Sb_{i})\leq P_i,~\forall i.
  \end{align}
\end{subequations}
Problem \eqref{eq: SRM k user} is NP-hard, but can be efficiently handled by Algorithm 1 (by letting $E_i=0$ $\forall i$) or existing block coordinate descent based methods \cite{RazaviHongLuo2012}.

{\bf 2) $K$-user TDMA (D) scheme:} The transmission interval is divided into $K$ time slots, each of which has a time fraction $\alpha_\ell\geq 0$, satisfying $\sum_{\ell=1}^K\alpha_\ell=1$; see Fig. \ref{fig: time splitting b} for the case of $K=3$. In the $\ell$th time slot, user $\ell$ operates in the ID mode; while the other $K-1$ users operate in the EH mode. Here we assume that transmitters operating in the EH mode send deterministic signals so that receivers operating in the ID mode can remove the cross-link signals (see Section \ref{subsec:TDMA_C}). 
Let $\Sb_{k\ell}$ be the signal covariance matrix employed by transmitter $i$ in the $\ell$th time slot, for $k,\ell=1,\ldots,K$. The design problem of this TDMA (D) scheme can be formulated as\vspace{-3mm}
\begin{subequations}\label{Problem::K:B-ID}
  \begin{align}
    \max_{\substack{(\alpha_1,\ldots,\alpha_K)\in \Omega,\Sb_{k\ell}\succeq\zerob\\k,\ell=1,\ldots,K}}~&\sum_{\ell=1}^K w_\ell\alpha_\ell\log_2\left(1+\frac{\hb_{\ell\ell}^H\Sb_{\ell\ell}\hb_{\ell\ell}}{\sigma_\ell^2}\right)\\[-2pt]
    \st~~&
    \sum_{\ell\neq i}^K\alpha_\ell \sum_{k=1}^K\hb_{ki}^H\Sb_{k\ell}\hb_{ki}\geq E_i,~\forall i,\label{Problem::K:B-ID C1}\\[-2pt]
        &\Tr(\Sb_{k\ell})\leq P_k,~\forall k,\ell,\\[-2.1em]\notag
  \end{align}
\end{subequations}
where $\Omega\!=\!\{\{\alpha_\ell\}_{\ell=1}^K\,|\,\alpha_\ell\!\in\![0,1],\sum_{\ell=1}^K\alpha_\ell\!\leq\!1\}$, and
\eqref{Problem::K:B-ID C1} denotes the energy harvesting constraints of all users. Note that in \eqref{Problem::K:B-ID} we not only optimize the signal covariance matrices in all time slots but also optimize the time fractions $\{\alpha_\ell\}$.

Problem \eqref{Problem::K:B-ID} can be reformulated as a convex problem. To show this, define\vspace{-1mm}
\begin{align}
  \Wb_{k\ell}=\alpha_\ell \Sb_{k\ell},~k,\ell=1,\ldots,K.
\end{align}
Then, \eqref{Problem::K:B-ID} can be rewritten as
\begin{subequations}\label{Problem::K:B-ID2}
  \begin{align}\!\!\!
    \max_{\substack{(\alpha_1,\ldots,\alpha_K)\in \Omega,\Wb_{k\ell}\succeq\zerob\\k,\ell=1,\ldots,K}}~&\sum_{\ell=1}^K w_\ell\alpha_\ell\log_2\left(1+\frac{\hb_{\ell\ell}^H\Wb_{\ell\ell}\hb_{\ell\ell}}{\alpha_\ell\sigma_\ell^2}\right)\\[-2pt]
    \st~~&
    \sum_{\ell\neq i}^K \sum_{k=1}^K\hb_{ki}^H\Wb_{k\ell}\hb_{ki}\geq E_i,~\forall i,\\[-2pt]
        &\Tr(\Wb_{k\ell})\leq \alpha_\ell P_k,~\forall k,\ell.
  \end{align}
\end{subequations}
In \eqref{Problem::K:B-ID2}, all the constraints are linear. Besides, the function $\alpha_\ell\log_2 \left(1+{\hb_{\ell\ell}^H\Wb_{\ell\ell}\hb_{\ell\ell}}/({\alpha_\ell\sigma_\ell^2})\right)$
is concave since it is the perspective of the concave function $\log_2\left(1+{\hb_{\ell\ell}^H\Wb_{\ell\ell}\hb_{\ell\ell}}/{\sigma_\ell^2}\right)$. Therefore, problem \eqref{Problem::K:B-ID} is a convex optimization problem.

\vspace{-2mm}
\subsection{Practical Scheme by Power Splitting}\label{subsec: power splitting}

\begin{figure}[!t]\centering
 \includegraphics[width=0.65\linewidth]{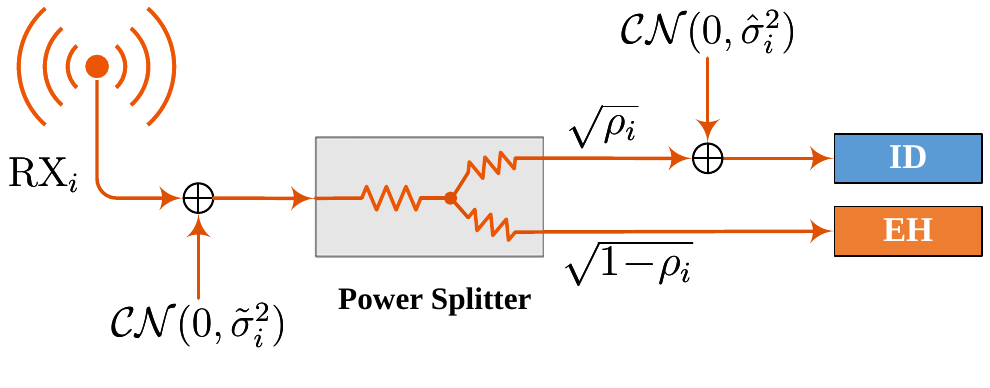}
 \caption{Diagram of the power splitting receiver for WIET.}
 \label{fig: power splitting}\vspace{-6mm}
\end{figure}

Other than the TDMS and TDMA schemes, another practical scheme, called {\it power splitting (PS)} \cite{Zhangrui2011}, splits the received signal into two parts for simultaneous EH and ID; see Fig. \ref{fig: power splitting}. In this subsection, we extend this scheme to the $K$-user interference channel. Specifically, suppose that receiver $i$ splits $\rho_i\in [0,1]$ fraction of power for ID and $1-\rho_i$ fraction of power for EH. The associated WIET design problem is given by
\begin{subequations}\label{Power splitting}
\begin{align}\!\!\!
\max_{\substack{\Sb_i\succeq\zerob,0\leq\rho_i\leq 1,\\i=1,\ldots,K}}
    & \sum_{i=1}^K \!w_i \log_2\!\left(\!1\!+\!\frac{\rho_i\hb_{ii}^H\Sb_i\hb_{ii}}{\rho_i\sum_{k\neq i}\hb_{ki}^H\Sb_k\hb_{ki}
\!+\! \rho_i \tilde \sigma_i^2 \!+\! \hat  \sigma_i^2}\right)\label{(PS)_a}\\
\st~&\sum_{k=1}^K \hb_{ki}^H\Sb_k\hb_{ki} \geq \frac{E_i}{1-\rho_i}~\forall i=1,\ldots,K,\label{(PS)_b}\\
    &\Tr(\Sb_i) \leq P_i~\forall i=1,\ldots,K,\label{(PS)_e}
\end{align}
\end{subequations}
where $\tilde \sigma_i^2$ denotes the noise power at the RF end while $\hat  \sigma_i^2$ denotes the processing noise power.
Note that, in problem \eqref{Power splitting}, we not only optimize the signal covariance matrices $\Sb_1,\ldots,\Sb_K$, but also the power splitting fractions $\rho_1,\ldots,\rho_K$ in the receivers.

Firstly, it is not difficult to infer from Proposition \ref{proposition:(P)_rnk1} that transmit beamforming is optimal to problem \eqref{Power splitting} as $K=2$. Secondly, for the general $K$-user case, we show that problem \eqref{Power splitting} can be efficiently handled in a manner similar to Algorithm 1. By introducing slack variables $\theta_i=1/\rho_i$, $i=1,\ldots,K$, one can write \eqref{Power splitting} as
\begin{subequations}\label{Power splitting 1}
\begin{align}\!\!\!
\max_{\substack{\Sb_i\succeq\zerob,\,0\leq\rho_i\leq 1,\\ \theta_i\geq 0, \\i=1,\ldots,K}}
    & \sum_{i=1}^K\! w_i \log_2\!\left(\!1\!+\!\frac{\hb_{ii}^H\Sb_i\hb_{ii}}{\sum_{k\neq i}\hb_{ki}^H\Sb_k\hb_{ki}
+ \theta_i\hat  \sigma_i^2 + \tilde \sigma_i^2}\right)\label{Power splitting 1_a}\\[-5pt]
\st~&\sum_{k=1}^K \hb_{ki}^H\Sb_k\hb_{ki} \geq \frac{E_i}{1-\rho_i},~i=1,\ldots,K,\label{Power splitting 1_b}\\
&  \theta_i \geq 1/\rho_i, ~i=1,\ldots,K,\label{Power splitting 1_c}\\
    &\Tr(\Sb_i) \leq P_i,~i=1,\ldots,K,\label{Power splitting 1_e}
\end{align}
\end{subequations}
where \eqref{Power splitting 1_c} would hold with equality at the optimum.
Note that both constraints \eqref{Power splitting 1_b} and \eqref{Power splitting 1_c} are convex. As a result, like problem \eqref{Problem::Kuser::RateMax}, the non-convexity of \eqref{Power splitting 1} is mainly due to the sum rate function.
Therefore, we can apply the log-exponential reformulation and SCA method in Section \ref{subsec: k user ideal} to \eqref{Power splitting 1}. In particular, like \eqref{Problem::Kuser::RateMax:approx}, at the $n$th iteration, one solves the following approximation problem
\ifconfver
\begin{subequations}\label{Problem::Kuser::RateMax:approx ps}
  \begin{align}
    &\{\Sb_1^\star[n],\ldots, \Sb_K^\star[n],\theta_1^\star[n],\ldots,\theta_K^\star[n]\}= \notag\\[-1pt]
    &\argmax_{{\Sb_i\succeq \zerob,x_i,y_i,\theta_i,\forall i=1,\ldots,K}}~~\sum_{i=1}^K w_i(x_i - y_i)\log_{2}e \label{Problem::Kuser::RateMax:approx ps:a}\\[-1pt]
    &\!\!\st\,
     \sum_{k=1}^K\hb_{ki}^H\Sb_k\hb_{ki} + \theta_i\hat  \sigma_i^2 + \tilde \sigma_i^2 \geq e^{x_i}~\forall i,\label{approx ps:b}\\[-1pt]
    &~~~\sum_{k\neq i}^K\!\hb_{ki}^H\Sb_k\hb_{ki}\!+\!\theta_i\hat\sigma_i^2\!+\!\tilde\sigma_i^2\leq e^{\bar y_i[n]}(y_i\!-\!\bar y_i[n]\!+\!1)~\forall i,\!\!\!\label{approx ps:c}\\[-2pt]
    &~~~~\eqref{Power splitting 1_b}, \eqref{Power splitting 1_c}~{\rm and}~\eqref{Power splitting 1_e},
  \end{align}
\end{subequations}
\else
\begin{subequations}\label{Problem::Kuser::RateMax:approx ps}
  \begin{align}
    &\{\Sb_1^\star[n],\ldots, \Sb_K^\star[n],\theta_1^\star[n],\ldots,\theta_K^\star[n]\}= \notag\\
    &~~~~\arg~\max_{\substack{\Sb_i\succeq \zerob,x_i,y_i,\theta_i,\\\forall i=1,\ldots,K}}~~\sum_{i=1}^K w_i(x_i - y_i)\log_{2}e \label{Problem::Kuser::RateMax:approx ps:a}\\
    &~~~~~~~~~~~~~~~~~~~\st~
     \sum_{k=1}^K\hb_{ki}^H\Sb_k\hb_{ki} + \theta_i\hat  \sigma_i^2 + \tilde \sigma_i^2 \geq e^{x_i},~\forall i,\label{approx ps:b}\\
    &~~~~~~~~~~~~~~~~~~~~~~~~ \sum_{k\neq i}^K\hb_{ki}^H\Sb_k\hb_{ki} + \theta_i\hat  \sigma_i^2 + \tilde \sigma_i^2 \leq e^{\bar y_i[n] }(y_i - \bar y_i[n]  + 1),~\forall i,\label{approx ps:c}\\
    &~~~~~~~~~~~~~~~~~~~~~~~ \eqref{Power splitting 1_b}, \eqref{Power splitting 1_c}~{\rm and}~\eqref{Power splitting 1_e},
  \end{align}
\end{subequations}
\fi
\noindent{}where
$\bar y_i[n] \!= \!\ln\!\left(\sum_{k\neq i}^K \hb_{ki}^H\Sb_k^\star[n\!-\!1]\hb_{ki} + \theta_i^\star[n\!-\!1]\hat  \sigma_i^2 + \tilde \sigma_i^2\right)$, $i=1,\ldots,K$.

\section{Simulation Results and Discussions}\label{sec: simulation}

In this section, simulation results are presented to examine the performance of the proposed WIET schemes. Throughout the simulations, we assumed that each transmitter has identical, unit power budget, i.e. $P\triangleq P_1=\cdots=P_K=1$, and that the receiver noise powers are the same and equal to $0.1$, i.e., $\sigma^2\triangleq \sigma_1^2=\cdots = \sigma_K^2=0.1$. The signal-to-noise ratio (SNR), defined as $\SNR\triangleq P/\sigma^2$, is thus equal to $10$ dB. The channel vectors $\{\hb_{ki}\}$ were randomly generated following the complex Gaussian distribution $\hb_{ki}\sim\CN(\zerob,\Qb_{ki})$, where the channel covariance matrices $\Qb_{ki}\succ\zerob$ were randomly generated. We normalized the maximum eigenvalue of $\Qb_{ii}$, i.e., $\lambda_{\max}(\Qb_{ii})$, to one for all $i$, and normalized $\lambda_{\max}(\Qb_{ki})$ to a value $\eta>0$ for all $k\neq i$, $i=1,\ldots,K$. The parameter $\eta$ thereby represents the relative cross-link channel power.
All the results presented in this section were obtained by averaging over 500 independent channel realizations. For Algorithm 1, the stopping criterion was set to
\begin{align*}
  \frac{{\rm Rate}[n]-{\rm Rate}[n-1]}{{\rm Rate}[n-1]}\leq 10^{-3},
\end{align*}
where {\rm Rate}$[n]$ denotes the achieved sum rate at iteration $n$.
The Matlab package {\tt CVX} \cite{cvx} was used to solve the convex approximation problems
\eqref{Problem::Kuser::RateMax:approx}, \eqref{Problem::K:B-ID2} and \eqref{Problem::Kuser::RateMax:approx ps}.

{\bf Example 1 (Impact of cross-link channel power):}
We investigate how the cross-link channel power (i.e., $\eta$) can affect the performance of the proposed WIET schemes in the interference channels. We first consider the feasibility rate, defined as the ratio of the total number of channel realizations for which the energy requirement $E \triangleq E_1=E_2$ can be satisfied to the 500 randomly generated channel realizations, of the the ideal scheme, TDMS, TDMA, and PS schemes.
Fig. \ref{fig2:a} shows the results for $K=2$, $N_t=4$ and $E\in \{1,3\}$.
Notice from \eqref{(P)}, \eqref{TDMA_A_EH} and \eqref{Power splitting 1} that the ideal scheme, TDMS and PS schemes intrinsically have the same feasibility rate. Therefore, in Fig. \ref{fig2:a}, only the results of TDMS and TDMA are displayed. One can observe that the feasibility rates of all schemes improves as $\eta$ increases. This is owing to the fact that the cross-link interference signals can benefit energy harvesting. We also observe that the TDMS scheme is more likely to be feasible than the TDMA scheme.

Fig. \ref{fig2:b} shows the average sum rate versus $\eta$ achieved by the five schemes under consideration. Note that whenever a scheme is infeasible, the achievable sum rate was set to zero. The results were obtained by averaging over 500 channel realizations. Firstly, one can see that all schemes have improved sum rates as $\eta$ increases. This is because, from Fig. \ref{fig2:a}, the larger $\eta$ is, the easier for the receivers to harvest the energy; all schemes can therefore allocate more time and power resources for information transfer as $\eta$ increases. Secondly, one observes that
the ideal scheme, TDMS and PS schemes all outperform the TDMA and TDMA (D) schemes. This is because, given $N_t=4$ and $K=2$, the cross-link interference can in general be well controlled, and thus these spectrum sharing schemes admit higher data throughput. Thirdly, one can observe from Fig. \ref{fig2:b} that, when $\eta\leq 2.2$, the PS scheme outperforms the TDMS scheme; whereas, when $\eta > 2.2$, the TDMS scheme can yield higher sum rate. This is due to the fact that, when $\eta$ is large, the TDMS scheme will spend only a negligible fraction of time in energy harvesting, and use most of the time in information transfer. Since the ID mode of the TDMS scheme is free from any energy harvesting constraint, it can yield higher sum rate than the PS scheme. In fact, when both $\eta$ and $E$ are large, the TDMS scheme may even outperform the ideal scheme, as illustrated in the next example.

\ifconfver
\begin{figure}
 \centering
 \subfloat[Feasibility rate vs. $\eta$, for $E\in\{1,3\}$.]
          {\label{fig2:a}\includegraphics[width=0.86\linewidth]{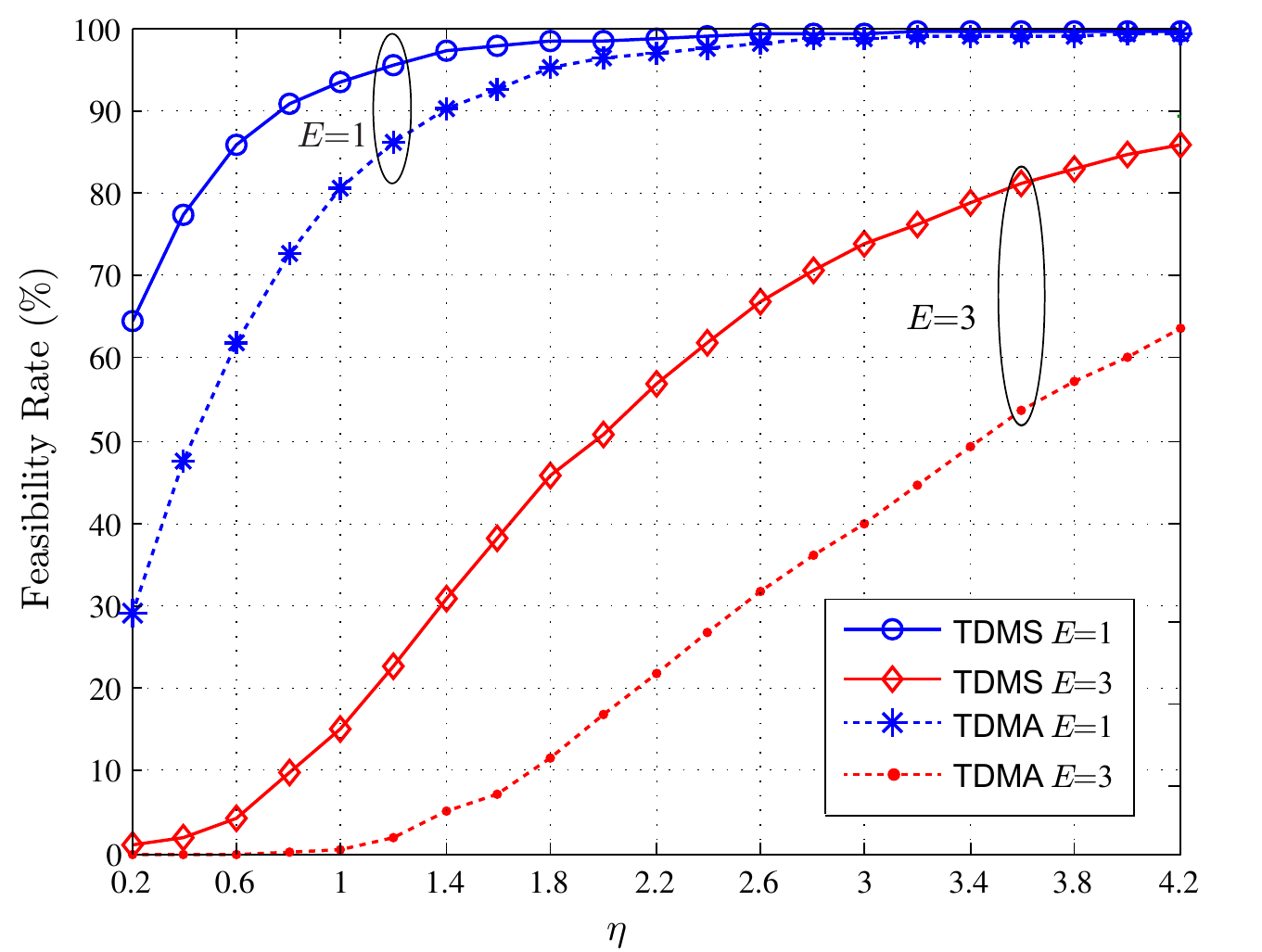}}\\
 \subfloat[Average sum rate vs $\eta$, for $E\!\!=\!\!1$.]
          {\label{fig2:b}\includegraphics[width=0.85\linewidth]{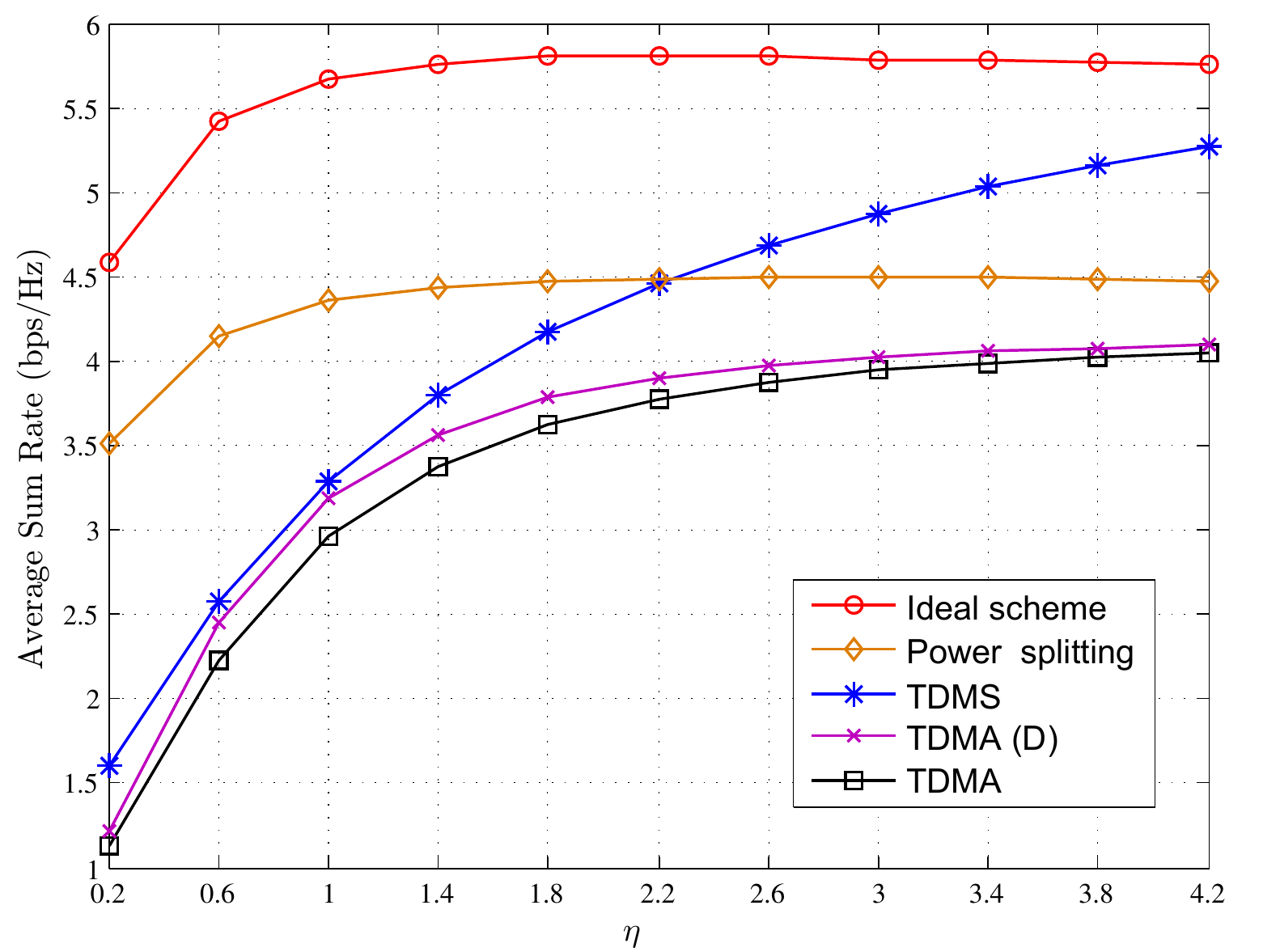}}
 \caption{Simulation results for the scenario with $K=2$, $N_t\!=\!4$ and ${\rm SNR}=10$ dB.}
 \label{fig2}\vspace{-0.5cm}
\end{figure}
\begin{figure}[!t]
 \centering
 \subfloat[Average sum rate vs. $E$, for $N_t=4$. ]{\label{fig3:a}
          \includegraphics[width=0.835\linewidth]{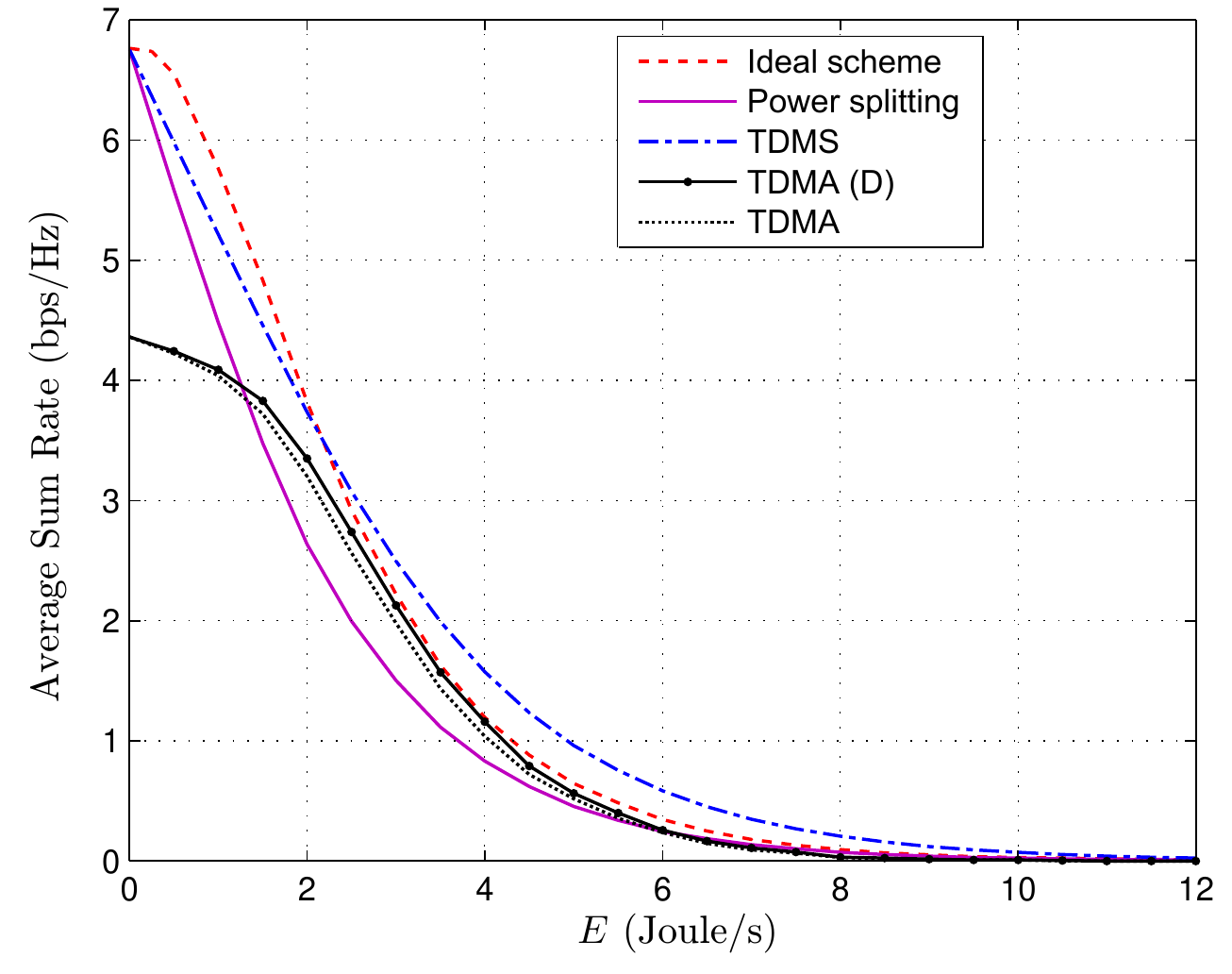}}\\
 \subfloat[Average sum rate vs. $E_2$, for $E_1=2$, $N_t=2$.]{\label{fig3:b}
          \includegraphics[width=0.82\linewidth]{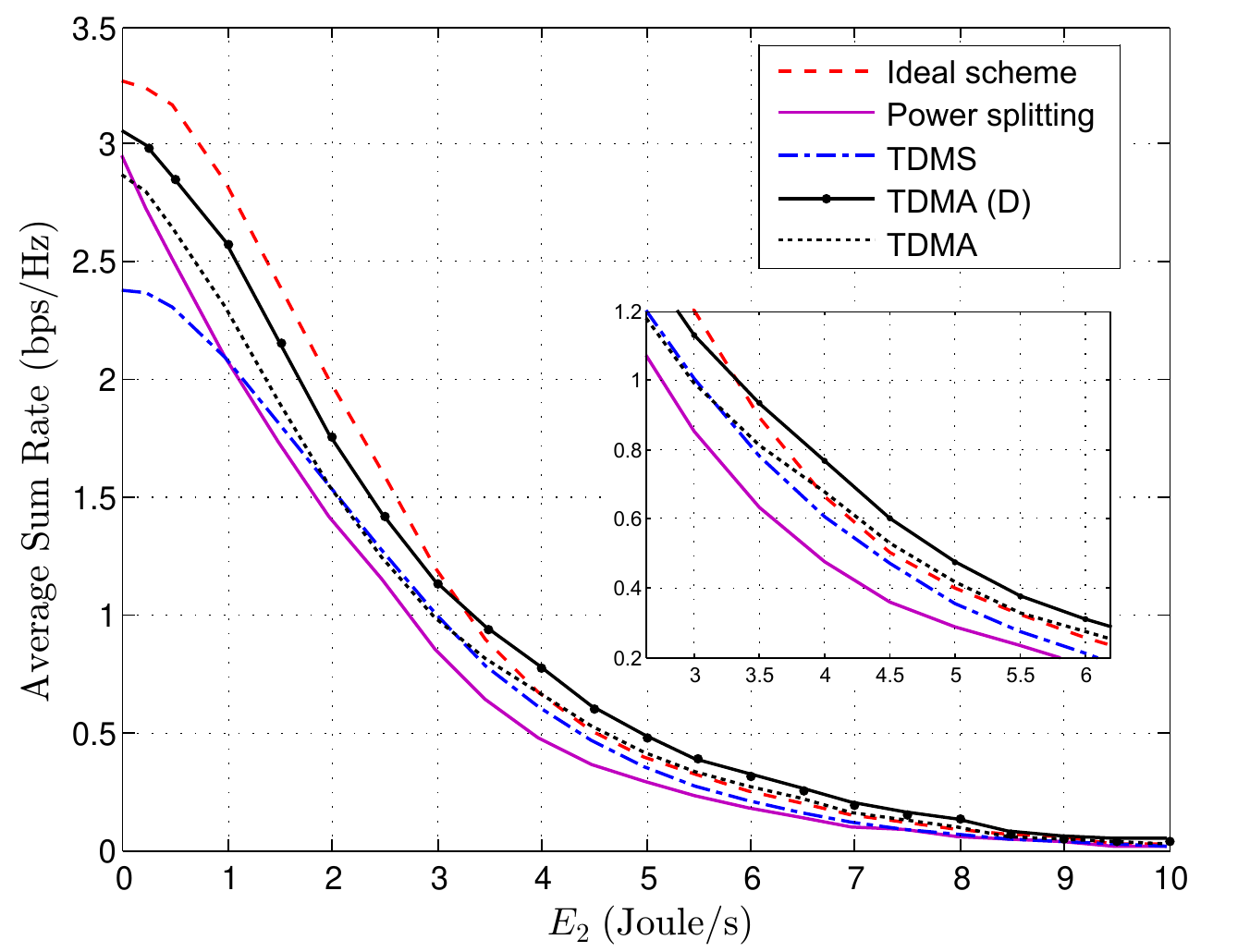}}
 \caption{Simulation results for the scenario with $K=2$, $\eta=4$ and $\SNR=10$ dB.}
 \label{fig3}\vspace{-0.5cm}
\end{figure}
\else
\begin{figure}[!t]
 \centering
 \subfloat[Feasibility rate vs. $\eta$, for $E\in\{1,3\}$.]
          {\label{fig2:a}\includegraphics[width=0.5\linewidth]{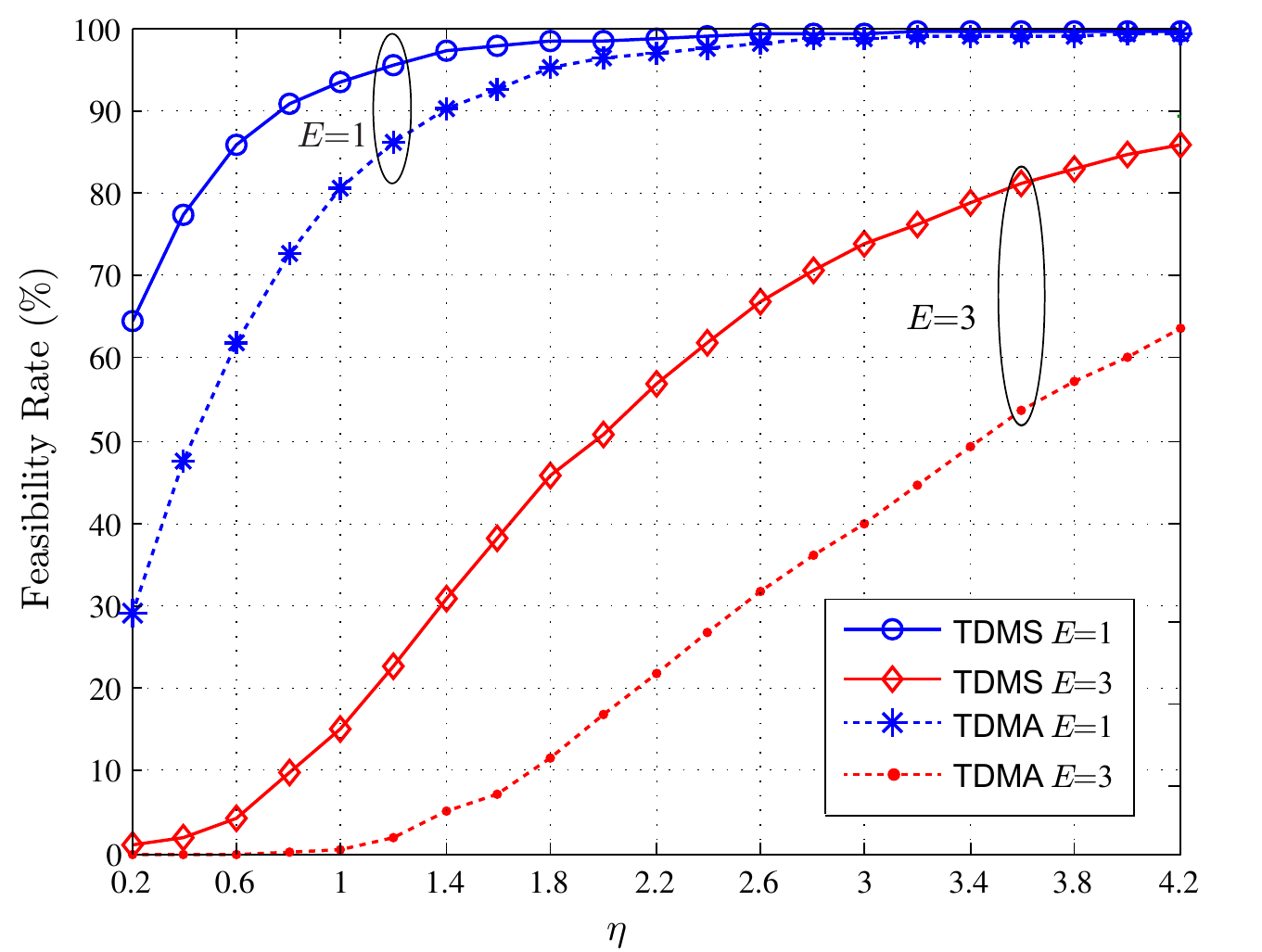}}~
 \subfloat[Average sum rate vs $\eta$, for $E\!\!=\!\!1$.]
          {\label{fig2:b}\includegraphics[width=0.5\linewidth]{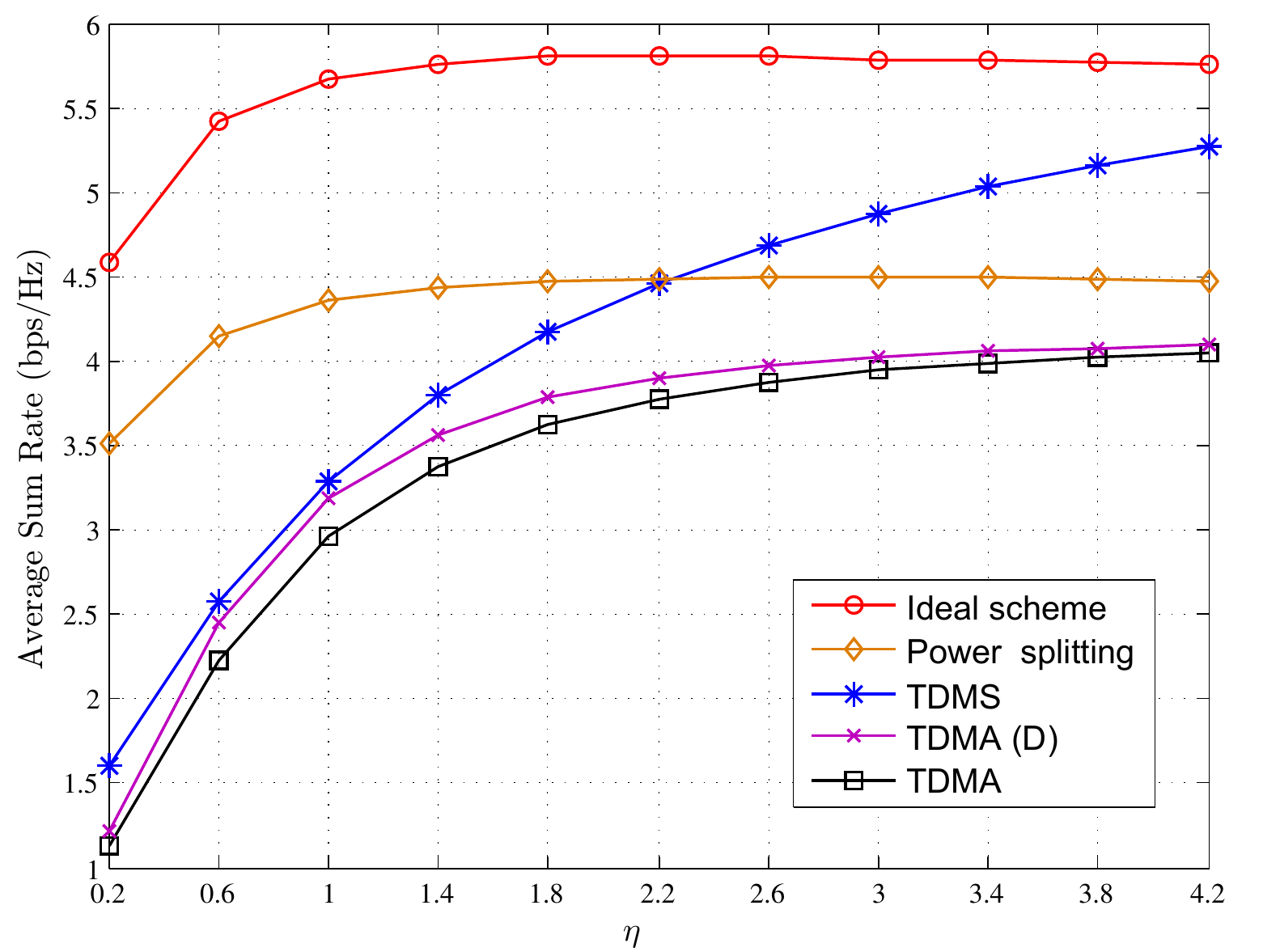}}
 \caption{Simulation results for the scenario with $K=2$, $N_t\!=\!4$ and ${\rm SNR}=10$ dB.}
 \label{fig2}\vspace{-0.5cm}
\end{figure}

\begin{figure}[!t]
 \centering
 \subfloat[Average sum rate vs. $E$, for $N_t=4$. ]{\label{fig3:a}\includegraphics[width=0.48\linewidth]{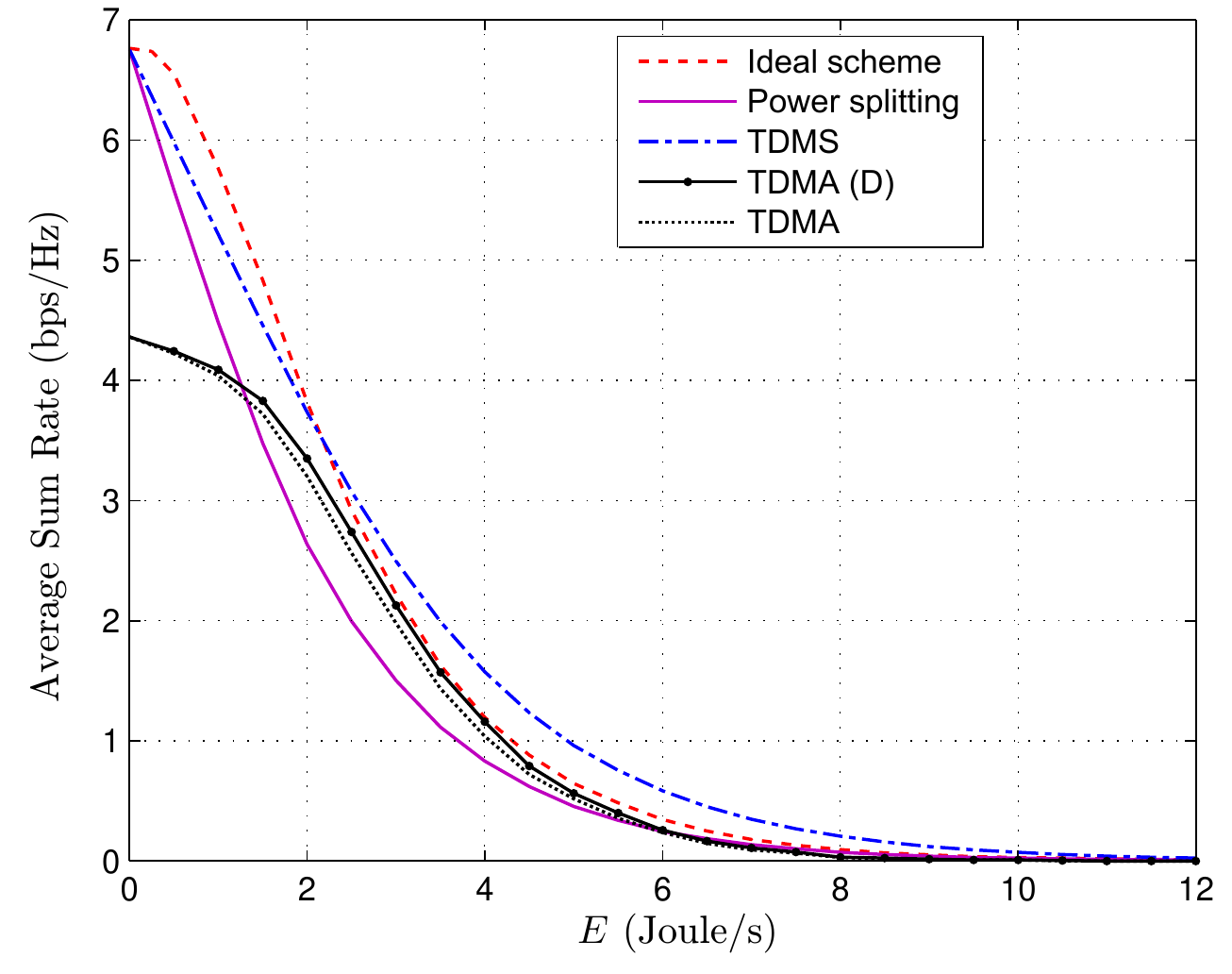}}
 \subfloat[Average sum rate vs. $E_2$, for $E_1=2$, $N_t=2$.]{\label{fig3:b}\includegraphics[width=0.5\linewidth]{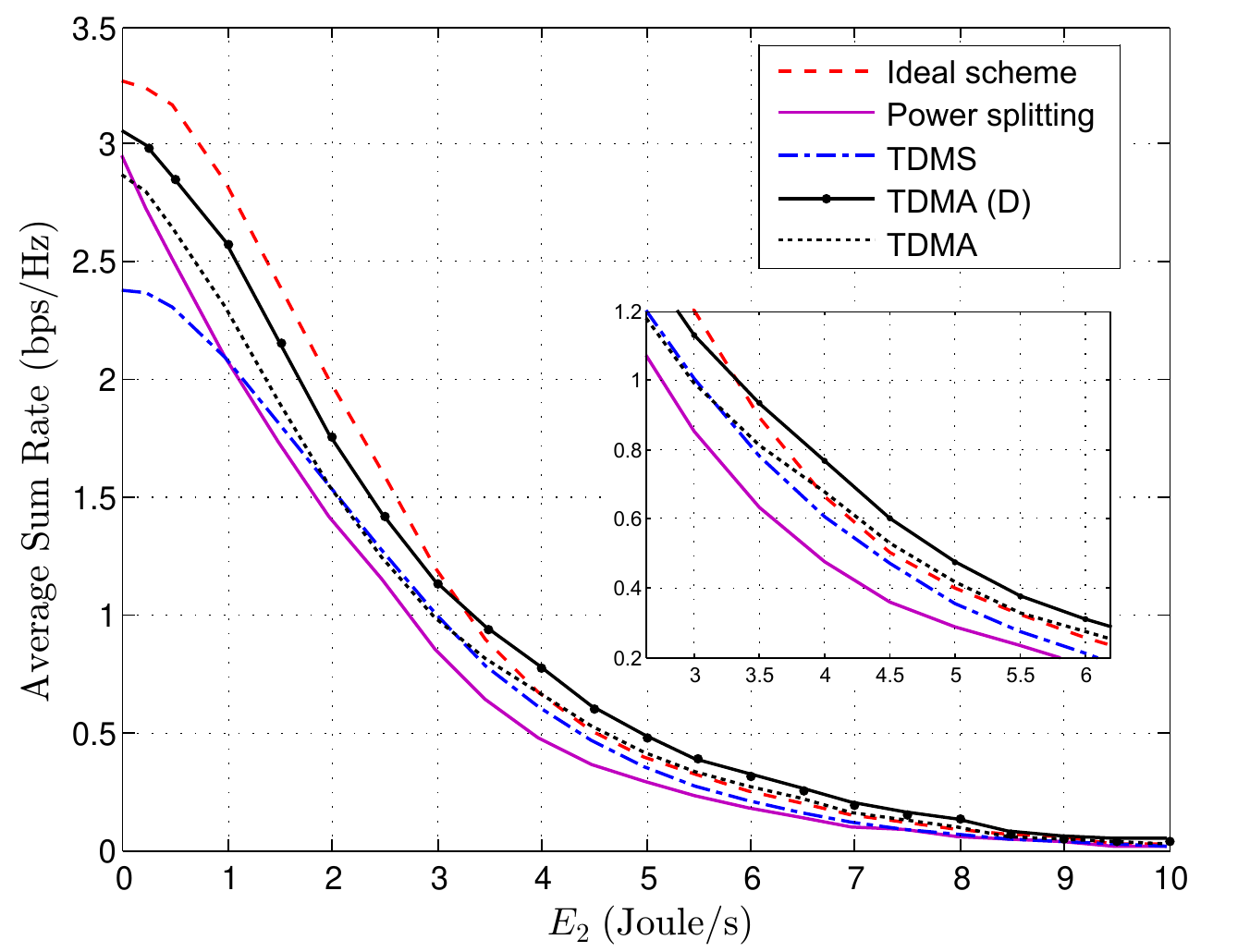}}
 \caption{Simulation results for the scenario with $K=2$, $\eta=4$ and $\SNR=10$ dB.}
 \label{fig3}\vspace{-1cm}
\end{figure}
\fi

{\bf Example 2 (Impact of the EH requirement):} Fig. \ref{fig3:a} shows the average sum rate versus the energy requirement $E \triangleq E_1=E_2$, for $N_t=4$ and $\eta=4$. As expected, the achievable sum rate decreases as the EH requirement increases. Moreover, when $E$ is small ($E \leq 2$), the ideal scheme can perform best; this is consistent with Property \ref{Property energy}. However, when $E>2$, the TDMS scheme outperforms the ideal scheme. It is also noted that when $E\geq 1.7$, the PS scheme exhibits the poorest sum rate performance.
In Fig. \ref{fig3:b}, we show the simulation results under an asymmetric energy requirement setting. In particular, we plot the average sum rate versus the energy requirement of receiver 2 $E_2$, given that the energy requirement of receiver 1 was fixed to 2 ($E_1=2$). Interestingly, we see from Fig. \ref{fig3:b} that when $E_2$ is large, the TDMA and TDMA (D) schemes can outperform the ideal scheme and perform best.

{\bf Example 3 (Performance for the $K$-user scenario):}
In this example, we consider an interference dominated scenario by setting $N_t=2$ and $K=4$.
Fig. \ref{fig4:a} displays the average sum rate versus $E$, for $\eta=1$. It can be observed from this figure that, except the ideal scheme, the TDMA (D) scheme outperforms the TDMS and PS schemes when $E\geq 1.3$. Fig. \ref{fig4:b} shows the simulation results for $\eta=4$. We observe that the TDMA (D) scheme instead yields highest sum rates when $E\geq 2$. Moreover, the TDMS scheme becomes to perform better than the ideal scheme and PS scheme when $E\geq 1.8$.

\ifconfver
\begin{figure}[!t]
 \centering
  \subfloat[Average sum rate vs. $E$, for $\eta=1.0$.]{\label{fig4:a}
            \includegraphics[width=0.86\linewidth]{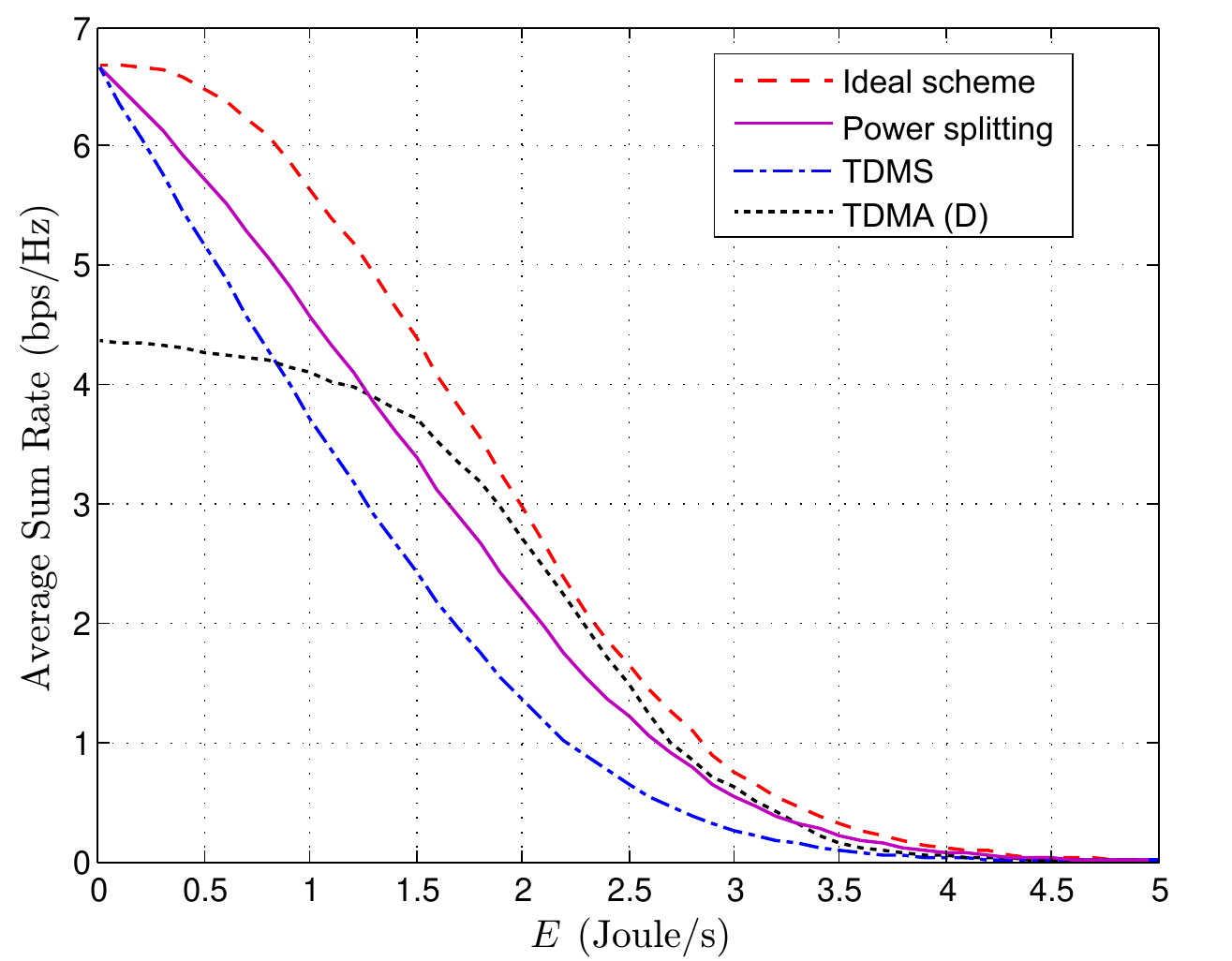}}\\
  \subfloat[Average sum rate vs. $E$, for $\eta=4.0$]{\label{fig4:b}\!\!\!\!
            \includegraphics[width=0.82\linewidth]{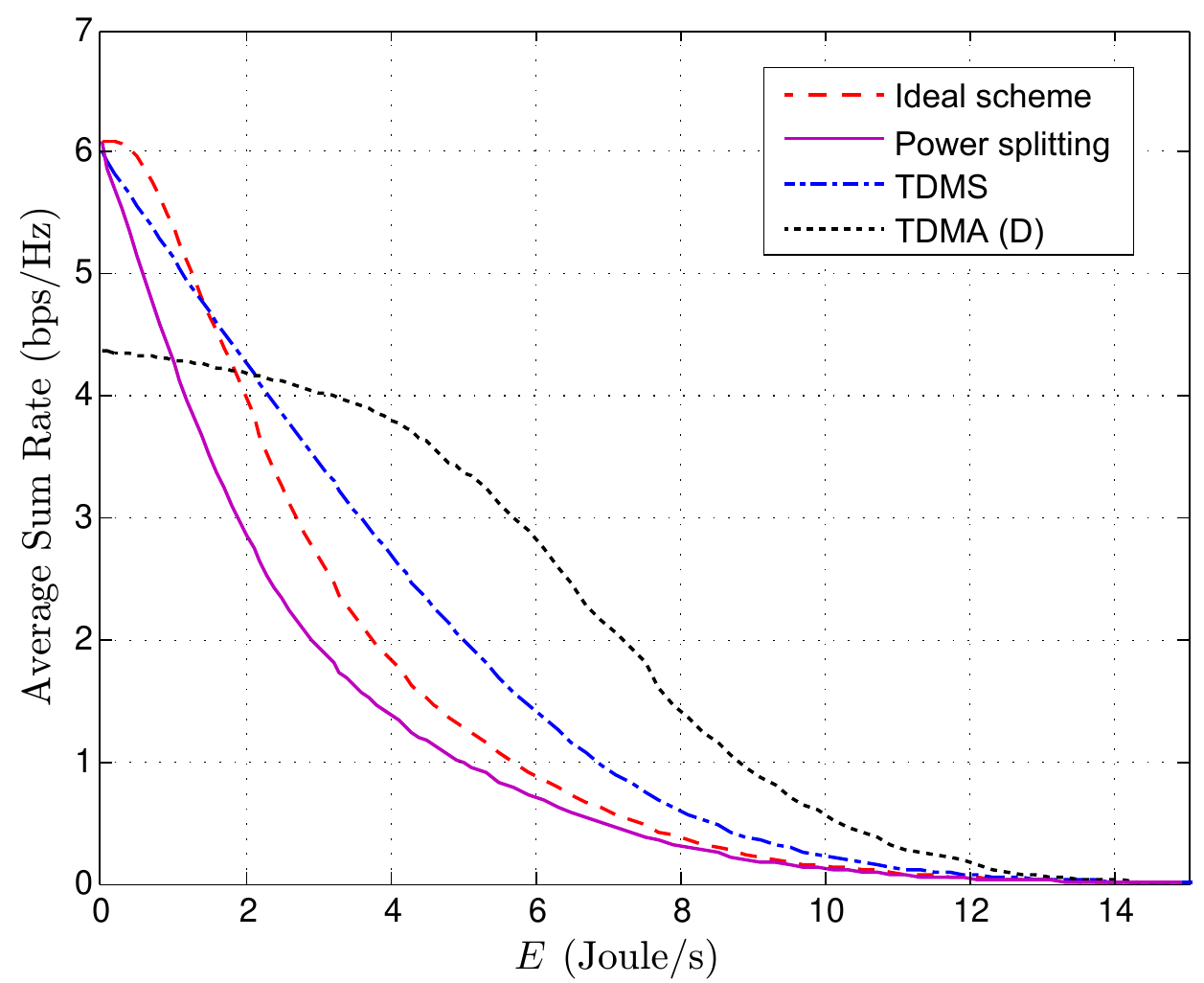}}
  \caption{Simulation results for the scenario with $K=4$, $N_t=2$ and $\SNR=10$ dB.}
  \label{fig4} \vspace{-0.5cm}
\end{figure}
\else
\begin{figure}[!t]
 \centering
  \subfloat[Average sum rate vs. $E$, for $\eta=1.0$.]{\label{fig4:a}\includegraphics[width=0.5\linewidth]
  {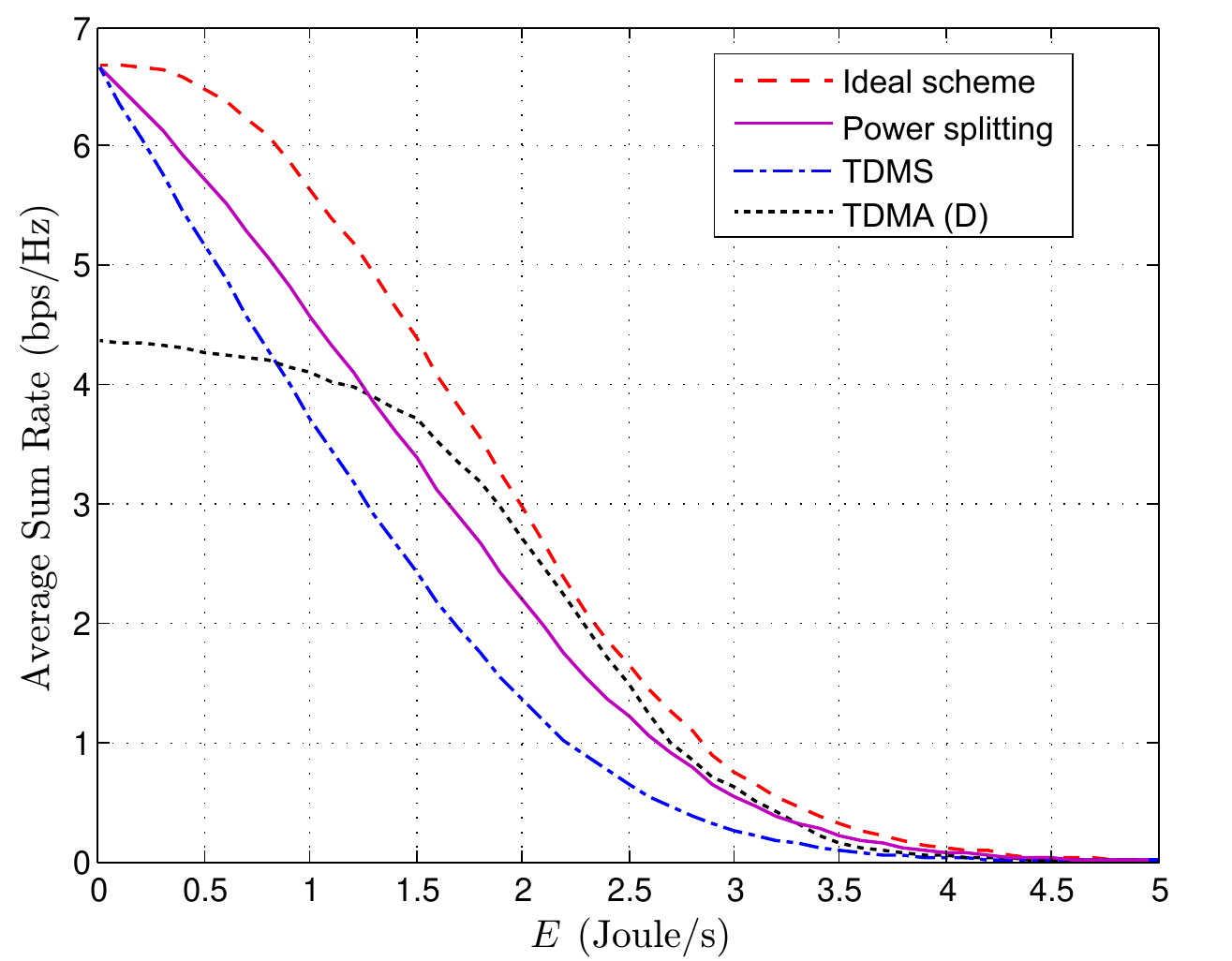}}~
  \subfloat[Average sum rate vs. $E$, for $\eta=4.0$]{\label{fig4:b}\includegraphics[width=0.48\linewidth]{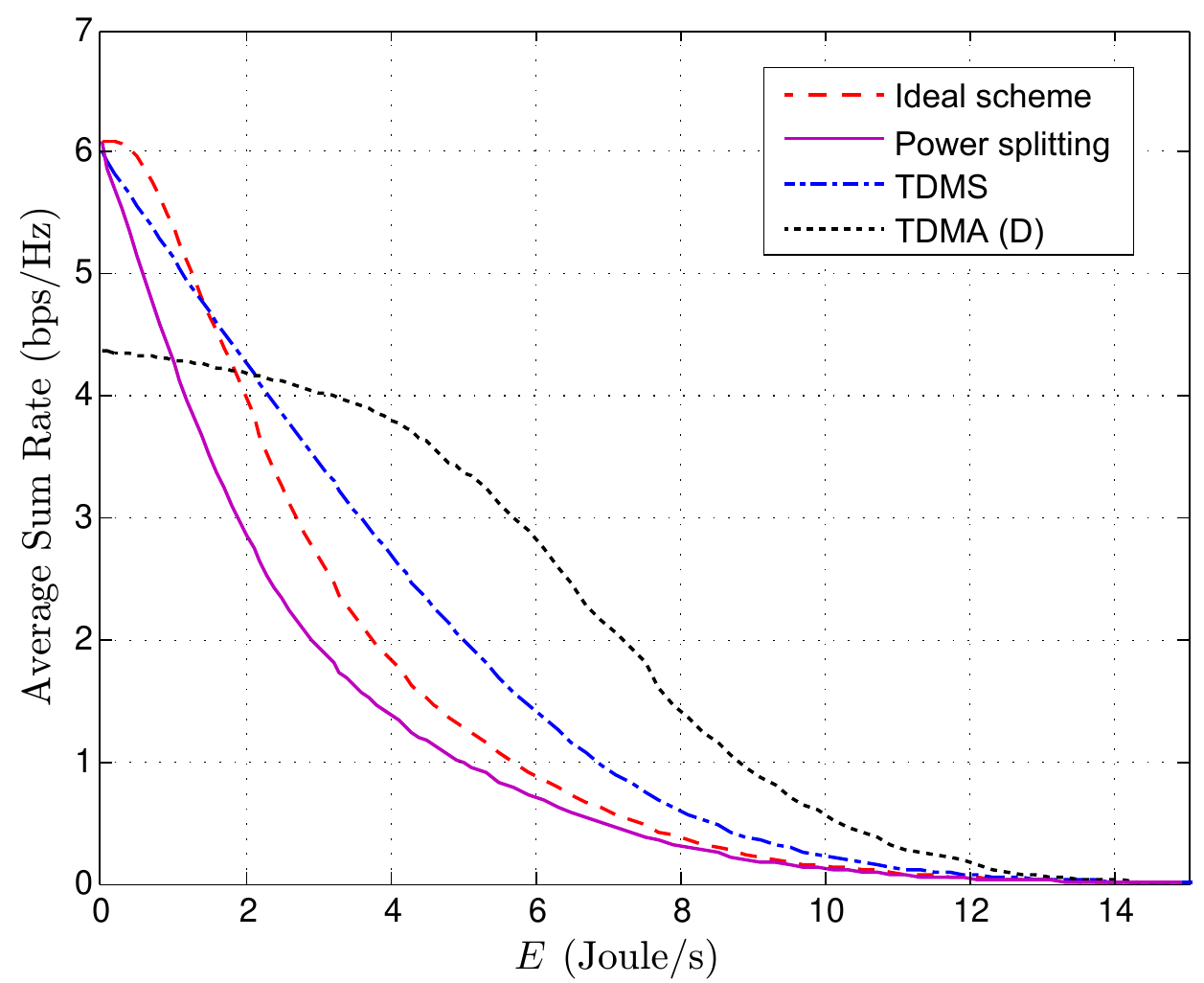}}~
  \caption{Simulation results for the scenario with $K=4$, $N_t=2$ and $\SNR=10$ dB.}
  \label{fig4} \vspace{-0.5cm}
\end{figure}
\fi

\section{Conclusions and Future Works}\label{sec: conclusions}
In this paper, we have considered the WIET problem in a multi-user MISO interference channel. In addition to the ideal scheme, we have proposed three practical schemes, namely, the TDMS, TDMA and PS schemes.
Starting with the two-user scenario, we have analyzed the optimal transmission strategy of the ideal scheme as well as semi-analytical solutions to the TDMS and TDMA schemes. It is shown that beamforming is optimal to these schemes. The proposed schemes have also been extended to the general $K$-user scenario. Specifically, we have shown that the design problems of the ideal scheme and the PS scheme can be efficiently handled by the proposed SCA method (Algorithm 1). The optimal transmit signal covariance matrices and optimal time fractions of the TDMA (D) scheme (energy harvesting using deterministic signals) can be obtained by solving a convex problem [i.e., \eqref{Problem::K:B-ID2}].

The simulation results have revealed interesting tradeoffs between EH and ID in the complex IFC. In particular, it has been observed that strong cross-link channel power is not detrimental under energy harvesting constraints; instead, the achievable sum rate can be improved with stronger cross-link channel powers. We have also observed that
none of the considered schemes can always dominate another in terms of the sum rate performance. For the three practical schemes, we have observed that
\begin{itemize}
\item when $N_t\geq K$, and $\eta$ and $E$ are not large, the PS scheme performs better than the TDMS and TDMA scheme on average;
\item when $N_t\geq K$, but $\eta$ and $E$ are large, the TDMS scheme in general performs best and can even outperform the ideal scheme {\sf (P)};
\item when $N_t<K$ and $E$ is large, the TDMA scheme in general can yield the highest sum rate.
\end{itemize}

The current work may motivate several interesting directions for future research. Firstly,
it is easy to see that, other than the considered $K$-user TDMS and TDMA schemes, there
exist other possible ways to separating the EH and ID modes of the $K$ receivers across the time. It would be interesting to see how the corresponding design problems can be efficiently solved and their performance compared to the schemes presented in this paper.
Secondly, since none of the considered schemes can always perform best, it is worth formulating a design formulation that unifies all these practical schemes. Thirdly, based on some insights gained from the current work, it is worthwhile to further study the WIET problems for some more complex interference channels, such as the broadcast interference channels \cite{ChaoTSP2012} and the MIMO interference channels \cite{Jafar2008}.

\vspace{-0.4cm}

\appendix
\setcounter{equation}{0}
\renewcommand{\theequation}{A.\arabic{equation}}

\subsection{Proof of Proposition 1}\label{appendix: proof of prop 1}

We prove by contradiction that $\Tr(\Sb_i^\star)=P_i$ for $i=1,2$. Suppose that $\Tr(\Sb_1^\star)<P_1$, then there exists some
$\epsilon>0$ and
\[ \Sb_1'=\Sb_1^\star+\epsilon\hat{\hb}_{12}^\perp (\hat{\hb}_{12}^\perp)^H \]
such that $\Tr(\Sb_1')=P_1$, where { $\hat\hb_{12}^\perp\triangleq \frac{\Pi_{\hb_{12}}^\perp\hb_{11}}{\|\Pi_{\hb_{12}}^\perp\hb_{11}\|}$}. Note that $(\Sb_1',\Sb_2^\star)$ is feasible to {\sf(P)}. Moreover, since $\hb_{11}\nparallel\hb_{12}$, we have $R_1(\Sb_1',\Sb_2^\star)>R_1(\Sb_1^\star,\Sb_2^\star)$ and $R_2(\Sb_1',\Sb_2^\star)=R_2(\Sb_1^\star,\Sb_2^\star)$, which contradicts the optimality of $(\Sb_1^\star,\Sb_2^\star)$. Hence, it must be that $\Tr(\Sb_1^\star)=P_1$; similarly, one can show that $\Tr(\Sb_2^\star)=P_2$.

Next, we show that $\Sb_1^\star$ and $\Sb_2^\star$ lie in the range space of $\Hb_1\triangleq[\hb_{11}~\hb_{12}]$ and $\Hb_2\triangleq[\hb_{21}~\hb_{22}]$, respectively, {i.e., $\Pi_{\Hb_i}^\perp \Sb_i^\star \Pi_{\Hb_i}^\perp=\zerob$ for $i=1,2$.} One can see that, for
any $\Sb\succeq\zerob$,
\begin{align} \label{eq: equivalence}
 \hb_{ik}^H(\Pi_{\Hb_i}\Sb\,\Pi_{\Hb_i}^H)\hb_{ik}&=\hb_{ik}^H\Sb\hb_{ik},\\
 \Tr(\Pi_{\Hb_i}\Sb\,\Pi_{\Hb_i}^H)&\le\Tr(\Sb),
\end{align}
for $i,k\!\in\!\{1,2\}$, where the equality in \eqref{eq: equivalence} holds because $\Pi_{\Xb}\Xb\!=\!\Xb$ for all $\Xb\!\in\!\Cset^{m\times{n}}$. Therefore, $(\Sb_1^\star,\Sb_2^\star)$ is an optimal solution to problem {\sf(P)} only if $(\Pi_{\Hb_1}\Sb_1^\star\Pi_{\Hb_1},\Pi_{\Hb_2}\Sb_2^\star\Pi_{\Hb_2})$ is optimal to {\sf(P)}. Now suppose that $\Sb_1^\star$ does not lie in the range space of $\Hb_1$, i.e., $\Tr(\Pi_{\Hb_1}^\perp\Sb_1^\star\Pi_{\Hb_1}^\perp)>0$. Then,
\[\Tr(\Pi_{\Hb_1}\Sb_1^\star\Pi_{\Hb_1}^H)\!=\!\Tr(\Sb_1^\star)-\Tr(\Pi_{\Hb_1}^\perp\Sb_1^\star\Pi_{\Hb_1}^\perp)\!<\!\Tr(\Sb_1^\star)\le{P_1},\]
which implies that $\Pi_{\Hb_1}\Sb_1^\star\Pi_{\Hb_1}$ is not optimal, and thereby $\Sb_1^\star$ is not optimal to {\sf(P)}. Analogously, one can show that $\Sb_2^\star$ must lie in the range space of $\Hb_2$.

What remains to prove \eqref{optimal S} is to show that there exists a pair of $(\Sb_1^\star,\Sb_2^\star)$ that are of rank one. It is not difficult to see that {\sf(P)} is equivalent to the following problems
\begin{subequations}\label{opt_Si}
\begin{align}
\max_{\Sb_i\succeq\zerob}~
    &\log\left(1+\frac{\hb_{ii}^H\Sb_i\hb_{ii}}{\Gamma_{ki}^\star + \sigma_i^2}\right)\\
\st~&\hb_{ik}^H\Sb_i\hb_{ik} + \Gamma_{kk}^\star \geq E_k,\\
    &\Gamma_{ki}^\star + \hb_{ii}^H\Sb_i\hb_{ii} \geq E_i,\\
    &\hb_{ik}^H\Sb_i\hb_{ik}\leq \Gamma_{ik}^\star,\\
    &\Tr(\Sb_i) \leq P_i,
\end{align}
\end{subequations}
where $\Gamma_{ki}^\star\!=\!\hb_{ki}^H\Sb_k^\star \hb_{ki}$, $i,k\!\in\!\{1,2\}$ and $i\!\neq\!k$. Let us focus on the case of $i=1$, $k=2$, and rewrite \eqref{opt_Si} as
\begin{subequations}\label{opt_S1}
\begin{align}
\max_{\Sb_1\succeq\zerob}~
    &{\hb_{11}^H\Sb_1\hb_{11}}\label{opt_S1_a}\\
\st~&\hb_{12}^H\Sb_1\hb_{12} \geq E_2-\Gamma_{22}^\star,\label{opt_S1_b}\\
    &\hb_{12}^H\Sb_1\hb_{12} \leq \Gamma_{12}^\star,\label{opt_S1_d}\\
    &\hb_{11}^H\Sb_1\hb_{11} \geq E_1-\Gamma_{21}^\star,\label{opt_S1_c}\\
    &\Tr(\Sb_1) \leq P_1.\label{opt_S1_e}
\end{align}
\end{subequations}

Suppose that $\Gamma_{12}^\star=E_2-\Gamma_{22}^\star$. Then \eqref{opt_S1_b} and \eqref{opt_S1_d} merges to one equality constraint. In that case, \eqref{opt_S1} has only three inequality constraints. According to \cite[Theorem 3.2]{Huang2010}, problem \eqref{opt_S1} then has an optimal solution ${\Sb}_1^\star$ such that $\rank({\Sb}_1^\star)\le1$. On the other hand, if $\Gamma_{12}^\star>E_2-\Gamma_{22}^\star$, then one of the two constrains \eqref{opt_S1_b} and \eqref{opt_S1_d} must be inactive for $\Sb_1^\star$. Therefore, the effective number of inequalities in \eqref{opt_S1} is again three. It then follows from \cite{Huang2010} that $\rank({\Sb}_1^\star)\le1$.
The above results imply that optimal $\Sb_1$ is of the form\vspace{-3pt}
\begin{align}
  \Sb_1^\star = (a_1\hb_{11}+b_1\hb_{12}) (a_1\hb_{11}+b_1\hb_{12})^H,
\end{align}
where $a_1, b_1 \in \Cset$.
Since any phase rotation of $a_1\hb_{11}+b_1\hb_{12}$ is invariant to $\Sb_1^\star$, we without loss of generality can let $a_1\in\Rset$.
Analogously, for the case of $i=2$, $k=1$, one can show that \eqref{opt_Si} has an optimal ${\Sb}_2^\star = (a_2\hb_{21}+b_2\hb_{22}) (a_2\hb_{21}+b_2\hb_{22})^H$, where $a_2\in\Rset$ and $b_1 \in \Cset$. The proof is thus complete. \hfill{$\blacksquare$}

\vspace{-3mm}
\subsection{Proof of Proposition \ref{proposition: one dimensional prob}}\label{appendix: proof of prop 2}
Firstly, note that problem \eqref{TDMA_A_EH} is equivalent to the max-main-fairness problem
\ifconfver
\begin{subequations}\label{TDMA_A_EH MMF}
\begin{align}\!
\max_{\Sb_1\succeq\zerob,\Sb_2\succeq\zerob}
&\min\left\{
  \frac{\sum_{i=1}^2\hb_{i1}^H\Sb_i\hb_{i1}}{E_1},
  \frac{\sum_{i=1}^2\hb_{i2}^H\Sb_i\hb_{i2}}{E_2}
 \right\}\label{TDMA_A_EH_MMF a}\\
\st&~~~\!\!\!\Tr(\Sb_1)\le{P_1},~\Tr(\Sb_2)\le{P_2}.\label{TDMA_A_EH_MMF d}
\end{align}
\end{subequations}
\else
\begin{subequations}\label{TDMA_A_EH MMF}
\begin{align}
\max_{\,\Sb_1\succeq\zerob,\,\Sb_2\succeq\zerob}~~&\min\left
\{\frac{\hb_{11}^H\Sb_1\hb_{11}+\hb_{21}^H\Sb_2
\hb_{21}}{E_1},~\frac{\hb_{12}^H\Sb_1\hb_{12}+\hb_{22}^H\Sb_2
\hb_{22}}{E_2}\right\}\label{TDMA_A_EH_MMF a}\\
\st&~~~\!\!\!\Tr(\Sb_1)\le{P_1},~\Tr(\Sb_2)\le{P_2}.\label{TDMA_A_EH_MMF d}
\end{align}
\end{subequations}\fi
Hence, given optimal $\Sb_1$ and $\Sb_2$, the optimal $\beta$ of \eqref{TDMA_A_EH} is given as in \eqref{solution of TDMA A EH 2}:
\begin{align}
\beta\!=\!\min\left
\{\frac{\hb_{11}^H\Sb_1\hb_{11}\!+\!\hb_{21}^H\Sb_2\hb_{21}}{E_1},\,
  \frac{\hb_{12}^H\Sb_1\hb_{12}\!+\!\hb_{22}^H\Sb_2\hb_{22}}{E_2}
\right\}.\label{beta sol}
\end{align}\vspace{-2mm}

Secondly, problem \eqref{TDMA_A_EH} satisfies the Slater's condition, so one can solve \eqref{TDMA_A_EH} by handling its Lagrange dual
problem. Let $\mu \geq 0$ and $\eta \geq 0$ be the Lagrange dual variables associated with constraints \eqref{TDMA_A_EH_b} and \eqref{TDMA_A_EH_c}, respectively. The dual problem of \eqref{TDMA_A_EH} can be shown as
\ifconfver
\begin{align}
\min_{\mu,\eta\ge0}&\!\left\{\!\!\!
\begin{array}{ll}
  \max\limits_{\Sb_1\succeq\zerob,\Sb_2\succeq\zerob}\!\!\!\!\!\!
  &\begin{pmatrix}
     \!\!\!\!\!\!\!\!\!\!\!\!\!\!\!\Tr(\Sb_1(\mu \hb_{11}\hb_{11}^H+\eta \hb_{12}\hb_{12}^H))\\
     ~~~~~+\Tr(\Sb_2(\eta\hb_{22}\hb_{22}^H+\mu  \hb_{21}\hb_{21}^H))
  \end{pmatrix}\\[8pt]
~~~~\st~&\Tr(\Sb_1)\le{P_1},~\Tr(\Sb_2)\le{P_2},
\end{array}\!\!\!\!
\right\}\notag \\
{\st}~&~ 1-E_1\mu-E_2\eta=0, \notag \\
&\!\!\!\!\!\!\!\!\!\!\!\!\!\!
=\min_{0\leq \mu \leq 1}\!\left\{\!\!\!
    \begin{array}{ll}
      \displaystyle\max_{\Sb_1\succeq\zerob,\Sb_2\succeq\zerob}\!\!\!\!&\tr(\Sb_1\Psib_1(\mu))\!+\!\tr(\Sb_2\Psib_2(\mu))\\
       ~~~~\st~&\Tr(\Sb_1 )\le{P_1},~\Tr(\Sb_2)\le{P_2},\label{TDMA_A_EH_dual}
    \end{array}\!\!\!\!\right\}\!\!
\end{align}
\else
\begin{align}
\min_{\substack{\mu,\eta \ge0}}~
&\left\{
\begin{array}{ll}
  \displaystyle\max_{\substack{\Sb_1\succeq\zerob,\Sb_2\succeq\zerob}}
\!\!\!\!&\tr(\Sb_1(\mu \hb_{11}\hb_{11}^H+\eta \hb_{12}\hb_{12}^H))
+\tr(\Sb_2(\eta\hb_{22}\hb_{22}^H+\mu  \hb_{21}\hb_{21}^H))
\\
~~~~~~~~~~~\st~&\Tr(\Sb_1)\le{P_1},~\Tr(\Sb_2)\le{P_2}.
\end{array}
\right\}
\notag \\
{\rm s.t.}~&~ 1-E_1\mu-E_2\eta=0, \notag \\
=\min_{\substack{0\leq \mu \leq 1}}~
&\left\{
\begin{array}{ll}
  \displaystyle\max_{\substack{\Sb_1\succeq\zerob,\Sb_2\succeq\zerob}}
\!\!\!\!&\tr(\Sb_1\Psib_1(\mu))
+\tr(\Sb_2\Psib_2(\mu))
\\
~~~~~~~~~~~\st~&\Tr(\Sb_1 )\le{P_1},~\Tr(\Sb_2)\le{P_2}.\label{TDMA_A_EH_dual}
\end{array}
\right\}
\end{align}
\fi
where $\Psib_1(\mu)=\mu \hb_{11}\hb_{11}^H+\frac{1-\mu E_1}{E_2} \hb_{12}\hb_{12}^H$ and
$\Psib_2(\mu)=\frac{1-\mu E_1}{E_2} \hb_{22}\hb_{22}^H+ \mu\hb_{21}\hb_{21}^H$. It is not difficult to show \cite[Proposition 2.1]{Zhangrui2011} that
\begin{align}\label{primal sol}
\Sb_1(\mu )=P_1\vb_1(\mu )\vb_1^H(\mu ),~\Sb_2(\mu )
=P_2\vb_2(\mu )\vb_2^H(\mu )
\end{align}
are optimal to the inner maximization problem of \eqref{TDMA_A_EH_dual}, where $\vb_i(\mu)\in \mathbb{C}^{N_t}$ is a principal eigenvector of $\Psib_i(\mu)$, for $i=1,2$.
As will be shown later, for $i=1,2$, under the assumption that $\hb_{i1}$ and $\hb_{i2}$ are linearly independent but not orthogonal to each other, $\Psib_i(\mu)$ has a unique maximum eigenvalue for any $\mu$. Hence, the solutions in \eqref{primal sol} are unique. According to the duality theory \cite{BK:BoydV04}, if $\mu$ is dual optimal (i.e., optimal to \eqref{TDMA_A_EH_dual}), then the unique $\Sb_1(\mu)$, $\Sb_2(\mu)$ in \eqref{primal sol} and $\beta$ in \eqref{beta sol} are optimal to problem \eqref{TDMA_A_EH}. The optimal $\mu$ can be obtained through a bisection search using the dual gradient, which is given by
\ifconfver
\begin{align*}
    &g =              \hb_{11}^H\Sb_1(\mu)\hb_{11}
        -\frac{E_1}{E_2}\hb_{22}^H\Sb_2(\mu)\hb_{22}\\
    &\hspace{26mm}
        -\frac{E_1}{E_2}\hb_{12}^H\Sb_1(\mu)\hb_{12}
                       +\hb_{21}^H\Sb_2(\mu)\hb_{21}.
\end{align*}
\else
$$ g =              \hb_{11}^H\Sb_1(\mu)\hb_{11}
    -\frac{E_1}{E_2}\hb_{22}^H\Sb_2(\mu)\hb_{22}
    -\frac{E_1}{E_2}\hb_{12}^H\Sb_1(\mu)\hb_{12}
                   +\hb_{21}^H\Sb_2(\mu)\hb_{21}.
$$
\fi

Lastly, we show that if $\hb_{i1}$ and $\hb_{i2}$ are linearly independent and $\Psib_i(\mu)$ has two equal eigenvalues, then $\hb_{i1}$ and $\hb_{i2}$ must be orthogonal. First note that $\range(\Psib_i(\mu)) =\range([\hb_{i1},\hb_{i2}])$ for linearly independent $\hb_{i1}$ and $\hb_{i2}$. Secondly, note that any principal eigenvector $\vb$ of $\Psib_i(\mu)$ belongs to  $\range(\Psib_i(\mu))$. If $\Psib_i(\mu)$ has two equal eigenvalues (the dimension of the principal eigenspace is two), then the principal eigenspace is exactly $\range([\hb_{i1},\hb_{i2}])$. Hence, $\tilde{\hb}_{i1}=\hb_{i1}/\|\hb_{i1}\|$,
$\tilde{\hb}_{i2}=\hb_{i2}/\|\hb_{i2}\|$,
$\tilde{\hb}_{i1}^\perp=\Pi_{\hb_{i1}}^\perp\hb_{i2}/\|\Pi_{\hb_{i1}}^\perp\hb_{i2}\|$
and
$\tilde{\hb}_{i2}^\perp=\Pi_{\hb_{i2}}^\perp\hb_{i1}/\|\Pi_{\hb_{i2}}^\perp\hb_{i1}\|$ are all principal eigenvectors of $\Psib_i(\mu)$. Let $\lambda_{\max}$ denote the principal eigenvalue of $\Psib_i(\mu)$, and now consider $i=1$. We have
\begin{subequations}\label{principal_eigvct}
\begin{align}
  \tilde{\hb}_{11}^H\Psib_1(\mu)\tilde{\hb}_{11}
=&\,\mu |\tilde{\hb}_{11}^H\hb_{11}|^2+\eta |\tilde{\hb}_{11}^H\hb_{12}|^2=\lambda,\!\label{principal_eigvct_a}\\
  \tilde{\hb}_{12}^H\Psib_1(\mu)\tilde{\hb}_{12}
=&\,\mu |\tilde{\hb}_{12}^H\hb_{11}|^2+\eta |\tilde{\hb}_{12}^H\hb_{12}|^2=\lambda,\!\label{principal_eigvct_b}\\
  (\tilde{\hb}_{11}^\perp)^H\Psib_1(\mu)\tilde{\hb}_{11}^\perp
=&\,\eta |(\tilde{\hb}_{11}^\perp)^H\hb_{12}|^2=\lambda,\label{principal_eigvct_c}\\
  (\tilde{\hb}_{12}^\perp)^H\Psib_1(\mu)\tilde{\hb}_{12}^\perp
=&\,\mu |(\tilde{\hb}_{12}^\perp)^H\hb_{11}|^2=\lambda,\label{principal_eigvct_d}
\end{align}
\end{subequations}
where $\eta=\frac{1-E_1\mu}{E_2}$. By \eqref{principal_eigvct_c} and \eqref{principal_eigvct_d}, we have
\[\eta |(\tilde{\hb}_{11}^\perp)^H\hb_{12}|^2=\mu |(\tilde{\hb}_{12}^\perp)^H\hb_{11}|^2.\]
Further combining \eqref{principal_eigvct_a} with \eqref{principal_eigvct_c} yields
\ifconfver
\begin{align}
&\mu |\tilde{\hb}_{11}^H\hb_{11}|^2+\eta |\tilde{\hb}_{11}^H\hb_{12}|^2=\eta |(\tilde{\hb}_{11}^\perp)^H\hb_{12}|^2,\notag\\
\Leftrightarrow~&\mu \|\hb_{11}\|^2+\eta (\|\hb_{12}\|^2-|(\tilde{\hb}_{11}^\perp)^H\hb_{12}|^2)=\eta |(\tilde{\hb}_{11}^\perp)^H\hb_{12}|^2,\notag\\
\Rightarrow~&\mu \|\hb_{11}\|^2+\eta \|\hb_{12}\|^2=2\eta |(\tilde{\hb}_{11}^\perp)^H\hb_{12}|^2\notag\\
=~&\mu |(\tilde{\hb}_{12}^\perp)^H\hb_{11}|^2+\eta |(\tilde{\hb}_{11}^\perp)^H\hb_{12}|^2.\label{equality}
\end{align}
\else
\begin{align}
&\mu |\tilde{\hb}_{11}^H\hb_{11}|^2+\eta |\tilde{\hb}_{11}^H\hb_{12}|^2=\eta |(\tilde{\hb}_{11}^\perp)^H\hb_{12}|^2,\notag\\
\Leftrightarrow~&\mu \|\hb_{11}\|^2+\eta (\|\hb_{12}\|^2-|(\tilde{\hb}_{11}^\perp)^H\hb_{12}|^2)=\eta |(\tilde{\hb}_{11}^\perp)^H\hb_{12}|^2,\notag\\
\Rightarrow~&\mu \|\hb_{11}\|^2+\eta \|\hb_{12}\|^2=2\eta |(\tilde{\hb}_{11}^\perp)^H\hb_{12}|^2=\mu |(\tilde{\hb}_{12}^\perp)^H\hb_{11}|^2+\eta |(\tilde{\hb}_{11}^\perp)^H\hb_{12}|^2.\label{equality}
\end{align}
\fi
Since both $\mu $, $\eta $ are nonnegative, and
$|(\tilde{\hb}_{12}^\perp)^H\hb_{11}|^2\le\|\hb_{11}\|^2$,
$|(\tilde{\hb}_{11}^\perp)^H\hb_{12}|^2\le\|\hb_{12}\|^2$, the
equality in \eqref{equality} implies that
$|(\tilde{\hb}_{12}^\perp)^H\hb_{11}|^2=\|\hb_{11}\|^2$ and
$|(\tilde{\hb}_{11}^\perp)^H\hb_{12}|^2=\|\hb_{12}\|^2$, i.e.,
$\hb_{11}$ and $\hb_{12}$ are orthogonal to each other. However, this contradicts the assumption that $\hb_{11}$ is not orthogonal to $\hb_{12}$. Hence, the principal eigenvector of $\Psib_1(\mu)$ is unique. Similarly, the principal eigenvector of $\Psib_2(\mu)$ can be shown unique. \hfill{$\blacksquare$}

\vspace{-3mm}
\subsection{Proof of Proposition \ref{proposition: TDMA B}}\label{appendix of proof of prop 3}

Problem \eqref{TDMA_B_1} is a quasi-convex problem. We first apply the idea of the Charnes-Cooper transformation \cite{Charnes1962} to recast
problem \eqref{TDMA_B_1} as a convex problem. To illustrate this, consider the following convex semidefinite program (SDP)
\begin{subequations}\label{CC_trans_2}
\begin{align}
\max_{\Xb_1\succeq\zerob,\,\Xb_2\succeq\zerob,\,y\ge0}~&\!\!\alpha\log_2\left(1+\hb_{11}^H\Xb_1\hb_{11}\right)\\
\st~~~~~~~~~&\hspace{-2mm}\hb_{21}^H\Xb_2\hb_{21}+y\sigma_1^2=1, \label{CC_trans_2 b}\\
&\hspace{-2mm}\hb_{12}^H\Xb_1\hb_{12}+\hb_{22}^H\Xb_2\hb_{22}\ge{yE_2/\alpha},\\
&\hspace{-2mm}\Tr(\Xb_1)\le{yP_1},~\Tr(\Xb_2)\le{yP_2}.
\end{align}
\end{subequations}
Note that the optimal $y^\star$ of \eqref{CC_trans_2} must be positive; otherwise we have $\Xb^\star_1=\Xb_2^\star=\zerob$ which violates \eqref{CC_trans_2 b}.
Moreover, consider the following correspondence:
\begin{subequations}\label{transform}
\begin{align}
 &y=1/(\hb_{21}^H\Sb_2\hb_{21}+\sigma_1^2)>0, \\
 &\Xb_1=y\Sb_1,~\Xb_2=y\Sb_2.
\end{align}
\end{subequations}
Then, one can show that $(\Sb_1,\Sb_2)$ is feasible to \eqref{TDMA_B_1} if and  only if $(\Xb_1,\Xb_2,y)$ is feasible to \eqref{CC_trans_2}. Furthermore, the objective value achieved by $(\Sb_1,\Sb_2)$ in \eqref{TDMA_B_1} is the same as the objective value achieved by $(\Xb_1,\Xb_2,y)$ in \eqref{CC_trans_2}. Therefore, the two problems \eqref{TDMA_B_1} and \eqref{CC_trans_2} are equivalent, and one can obtain $(\Sb_1^\star,\Sb_2^\star)$ of \eqref{TDMA_B_1} by solving the convex problem \eqref{CC_trans_2}.

To show how problem \eqref{CC_trans_2} can be efficiently solved, we rewrite \eqref{CC_trans_2} as
\begin{subequations}\label{appx prop 3 prob1}
  \begin{align}
  \max_{\Xb_1\succeq\zerob,\Xb_2\succeq\zerob,y\geq 0}&~\hb_{11}^H\Xb_1\hb_{11}\\
  \st~~& \hb_{21}^H\Xb_2\hb_{21} + y\sigma_1^2 \leq 1,\label{appx prop 3 prob1:b}\\
      & \hb_{12}^H\Xb_1\hb_{12} + \hb_{22}^H\Xb_2\hb_{22} \geq y\frac{E_2}{\alpha},\\
      & \Tr(\Xb_1)\leq y P_1,~\Tr(\Xb_2)\leq y P_2,
\end{align}
\end{subequations}
where the inequality constraint \eqref{appx prop 3 prob1:b} holds with equality at the optimum. The variable $y$ has a feasible region of $0\leq y \leq 1/\sigma_1^2$. We assume that a feasible $y$ is given and investigate the associated optimal $\Xb_1$ and $\Xb_2$ of problem \eqref{appx prop 3 prob1}, which are denoted by $\Xb_1(y)$ and $\Xb_2(y)$, respectively. One key observation is that $\Xb_2(y)$ can be obtained by solving the following problem
\begin{subequations}\label{appx prop 3 prob2}
  \begin{align}
  \Xb_2(y)=\arg~\max_{\Xb_2\succeq\zerob}&~~\hb_{22}^H\Xb_2\hb_{22}\\[-2pt]
  \st~ &~\hb_{21}^H\Xb_2\hb_{21} \leq 1- y\sigma_1^2,\label{appx prop 3 prob2:b}\\[-2pt]
       &~\Tr(\Xb_2)\leq y P_2.
\end{align}
\end{subequations}

Following \cite[Proposition 2.1]{Zhangrui2011}, problem \eqref{appx prop 3 prob2} has a closed-form solution as
\ifconfver
\begin{align}
\Xb_2(y)&\!=\!\vb_2(y)\vb_2^H(y),\label{X2(y)} \\
\vb_2(y)&\!=\!\left\{\!\!\!\!\!\!\!\!\!
        \begin{array}{ll}
            &\sqrt{y P_2}\bar\hb_{22},~~~{\rm if~} y P_2 |\hb_{21}^H\bar\hb_{22}|^2 < 1-y\sigma_1^2,\\[4pt]
            &\!\!
            \begin{pmatrix}
              \hspace{-10mm}\frac{\sqrt{1- y\sigma_1^2}}{|\hb_{21}^H\hb_{22}|}(\bar\hb_{21}^H\hb_{22})\bar\hb_{21}\\
              \hspace{ 8mm}+\sqrt{y P_2-\frac{1- y\sigma_1^2}{\|\hb_{21}\|^2}}\bar\hb_{21}^\perp
            \end{pmatrix}\!,
              ~{\rm otherwise.}
        \end{array}
        \right. \!\!\!\!\!\!\!\!\!\label{v2(y)}
\end{align}
\else
\begin{align}
\Xb_2(y)&=\vb_2(y)\vb_2^H(y),\label{X2(y)} \\
\vb_2(y)&=\left\{\!\!\!\!\!\!\!\!\!
\begin{array}{ll}
 &\sqrt{y P_2}\bar\hb_{22},~~~~~~~~~~~~~~~\,~~~~~{\rm if~} y P_2 |\hb_{21}^H\bar\hb_{22}|^2 < 1-y\sigma_1^2,\\
 &\frac{\sqrt{1- y\sigma_1^2}}{|\hb_{21}^H\hb_{22}|}(\bar\hb_{21}^H\hb_{22})\bar\hb_{21}
 + \sqrt{y P_2-\frac{1- y\sigma_1^2}{\|\hb_{21}\|^2}}\bar\hb_{21}^\perp,~  {\rm otherwise.}
\end{array}
\right. \label{v2(y)}
\end{align}
\fi

Notice that, if $y P_2|\hb_{21}^H\bar\hb_{22}|^2 \!<\! 1\!-y\sigma_1^2$, then $\hb_{21}^H\Xb_2(y)\hb_{21} \!<\! 1\!-\! y\sigma_1^2$, and thus $(\Xb_2(y),y)$ won't be optimal to problem \eqref{appx prop 3 prob1} since \eqref{appx prop 3 prob1:b} should hold with equality at the optimum.
Therefore, we can focus on the case of $\frac{1}{P_2|\hb_{21}^H\bar\hb_{22}|^2 + \sigma_1^2}\!\leq\! y \!\leq\! 1/\sigma_1^2$.
Let $g(y) \triangleq \hb_{22}^H\Xb_2(y)\hb_{22}=|\hb_{22}^H\vb_2(y)|^2$.
Given $\frac{1}{P_2|\hb_{21}^H\bar\hb_{22}|^2 + \sigma_1^2}\leq y \leq 1/\sigma_1^2$ and $\Xb_2(y)$, \eqref{appx prop 3 prob1}  reduces to
\begin{subequations}\label{Problem:A15}
  \begin{align}
    \Xb_1(y)=\argmax_{\Xb_1\succeq\zerob} ~&\hb_{11}^H\Xb_1\hb_{11}\\[-2pt]
    \st~~&\hb_{12}^H\Xb_1\hb_{12}\geq y\frac{E_2}{\alpha} - g(y),\\
         &\Tr(\Xb_1) \leq yP_1.
  \end{align}
\end{subequations}
Again, using \cite[Proposition 2.1]{Zhangrui2011}, problem \eqref{Problem:A15} has the optimal solution given by
\ifconfver
\begin{align}
\Xb_1(y)&\!=\!\vb_1(y)\vb_1^H(y),\label{X1(y)} \\
\vb_1(y)&\!=\!\left\{\!\!\!\!\!\!\!\!
\begin{array}{ll}
 &{\rm infeasible,}~~~~~~~~{\rm if~}y\frac{E_2}{\alpha} - g(y)> yP_1\|\hb_{12}\|^2,\\[3pt]
 &\sqrt{yP_1}\bar\hb_{11},~~~~~~~~{\rm if~} y\frac{E_2}{\alpha} - g(y)\leq yP_1|\bar\hb_{11}^H\hb_{12}|^2,\\[3pt]
 &
 \begin{pmatrix}
   \hspace{-10mm}\frac{\sqrt{yE_2/\alpha-g(y)}}{|\hb_{12}^H\hb_{11}|}\bar\hb_{12}^H\hb_{11}\bar\hb_{12}\\[3pt]
   \hspace{ 8mm}+\sqrt{yP_1\!-\!\frac{y{E_2}/{\alpha} - g(y)}{\|\hb_{12}\|^2}}\bar\hb_{12}^\perp
 \end{pmatrix},~{\rm otherwise.}
\end{array}
\right. \notag\label{v1(y)}
\end{align}
\else
\begin{align}
\Xb_1(y)&=\vb_1(y)\vb_1^H(y),\label{X1(y)} \\
\vb_1(y)&=\left\{\!\!\!\!\!\!\!\!
\begin{array}{ll}
 &{\rm infeasible,}~~~~~~~~~~~~~~~~~~~~~~~~~~{\rm if~}y\frac{E_2}{\alpha} - g(y)> yP_1\|\hb_{12}\|^2,\\[3pt]
 &\sqrt{yP_1}\bar\hb_{11},~~~~~~~~~~~~~~~~~~~~~~~~~~~{\rm if~} y\frac{E_2}{\alpha} - g(y)\leq yP_1|\bar\hb_{11}^H\hb_{12}|^2,\\[3pt]
 &\frac{\sqrt{y{E_2}/{\alpha} - g(y)}}{|\hb_{12}^H\hb_{11}|}\bar\hb_{12}^H\hb_{11}\bar\hb_{12}+ \sqrt{yP_1\!-\!\frac{y{E_2}/{\alpha} - g(y)}{\|\hb_{12}\|^2}}\bar\hb_{12}^\perp,~~~  {\rm otherwise.}
\end{array}
\right. \label{v1(y)}
\end{align}
\fi

Therefore, given a $\frac{1}{P_2|\hb_{21}^H\bar\hb_{22}|^2 + \sigma_1^2}\leq y \leq 1/\sigma_1^2$, one can efficiently obtain $\Xb_1(y)$ and $\Xb_2(y)$ by \eqref{X1(y)} and \eqref{X2(y)}, respectively.
The optimal $y$ of problem \eqref{appx prop 3 prob1} then can be obtained by solving the following one-dimensional problem
\begin{subequations}\label{Eq:optimal y1}
  \begin{align}
  y^\star= \argmax_{y}~&\hb_{11}^H\Xb_1(y)\hb_{11} \\[-5pt]
  \st~~~&\frac{1}{P_2|\hb_{21}^H\bar\hb_{22}|^2+\sigma_1^2} \leq y \leq {  \frac{1}{\sigma_1^2}}.
 \end{align}
\end{subequations}
The function $\hb_{11}^H\Xb_1(y)\hb_{11}$ is in fact concave in $y$, and hence \eqref{Eq:optimal y1} can be solved via bisection. To show this, note from \eqref{appx prop 3 prob1} that
\begin{subequations}\label{TDMA_B_y}
\begin{align}\hspace{-3mm}
\hb_{11}^H\Xb_1(y)\hb_{11}\!=\!\max_{\Xb_1\succeq\zerob,\Xb_2\succeq\zerob}&~\hb_{11}^H\Xb_1\hb_{11}\\
\st\hspace{13mm}
&\hspace{-11mm}\hb_{21}^H\Xb_2\hb_{21}+y\sigma_1^2\le1,\\
&\hspace{-11mm}\hb_{12}^H\Xb_1\hb_{12}+\hb_{22}^H\Xb_2\hb_{22}\ge{y}\frac{E_2}{\alpha},\\
&\hspace{-11mm}\Tr(\Xb_1)\le{yP_1},~\Tr(\Xb_2)\le{yP_2}.
\end{align}
\end{subequations}
Since problem \eqref{appx prop 3 prob1} is convex jointly in $(\Xb_1,\Xb_2,y)$, and $\hb_{11}^H\Xb_1(y)\hb_{11}$ is a ``point-wise" maximum of the jointly concave (linear) $\hb_{11}^H\Xb_1\hb_{11}$ over all $(\Xb_1,\Xb_2)$ feasible to \eqref{TDMA_B_y}. By \cite{BK:BoydV04}, $\hb_{11}^H\Xb_1(y)\hb_{11}$ is concave in $y$. The proof is thus complete. \hfill{$\blacksquare$}

\vspace{-3mm}
\subsection{Proof of Proposition \ref{proposition: conv of sca}}\label{Appendix proof of conv of sca}

We show that Algorithm 1 essentially belongs to the SUM method in \cite{RazaviHongLuo2012}. Note that, at the optimum, the inequalities in \eqref{approx:b} and \eqref{approx:c} of problem \eqref{Problem::Kuser::RateMax:approx} will hold with equality. Therefore,
problem \eqref{Problem::Kuser::RateMax:approx} can be equivalently expressed as\vspace{-2mm}
\begin{subequations}\label{Prob::Kuser::RateMax:approx}
  \begin{align}\!\!
    \{\Sb_i^\star[n]\}_{k=1}^K\!=\! \argmax_{\{\Sb_i\succeq \zerob\}_{i=1}^{K}}~
    &U(\Sb_1,\ldots, \Sb_K\,|\, \{\bar y_i[n]\}_{i=1}^K)\!\! \label{Problem::Kuser::RateMax:approx:a}\\[-2pt]
    \st~~&\eqref{Problem::Kuser::RateMax:c},~\eqref{Problem::Kuser::RateMax:d},\\[-22pt]\notag
  \end{align}
\end{subequations}
where
\ifconfver
\begin{align*}
  &U\left(\Sb_1,\ldots, \Sb_K~|~ \{\bar y_i[n]\}_{i=1}^K\right)\\
  &\!\!\!=\!\sum_{i=1}^K\!w_i\log_2\!\!
  \left(\!
        \frac{\sum_{i=1}^K\hb_{ik}^H\Sb_i\hb_{ik} + \sigma_i^2}
             {\exp{\!\left[\left(\!\sum_{k\neq i}\hb_{ki}^H\Sb_k\hb_{ki} \!+\! \sigma_i^2\right)\!e^{-\bar y_i[n]} \!+\!\bar y_i[n] \!-\!1 \right]}}
  \!\right).
\end{align*}
\else
\begin{align*}
  U\left(\Sb_1,\ldots, \Sb_K~|~ \{\bar y_i[n]\}_{i=1}^K\right)=
  \sum_{i=1}^K w_i\log_2
  \left(
        \frac{\sum_{i=1}^K\hb_{ik}^H\Sb_i\hb_{ik} + \sigma_i^2}
             {\exp{\left[\left(\sum_{k\neq i}\hb_{ki}^H\Sb_k\hb_{ki} + \sigma_i^2\right) e^{-\bar y_i[n]} +\bar y_i[n] -1 \right]}}
  \right).
\end{align*}
\fi
By the fact of
\ifconfver
$e^{y_i}\!\geq\! e^{\bar y_i[n]}(y_i\!-\!\bar y_i[n] \!+\! 1)~\forall y_i
\Leftrightarrow e^{ (e^{y_i}) e^{-\bar y_i[n]} +\bar y_i[n] -1}\!\geq\! e^{y_i}~\forall y_i$,
\else
\begin{align}
   e^{y_i}\geq e^{\bar y_i[n]}(y_i-\bar y_i[n] + 1)~\forall y_i
   \Longleftrightarrow  e^{ (e^{y_i}) e^{-\bar y_i[n]} +\bar y_i[n] -1 } \geq e^{y_i}~\forall y_i,
\end{align}
\fi
we see that $\exp\left(\big(\sum_{k\neq i}\hb_{ki}^H\Sb_k\hb_{ki} + \sigma_i^2\big) e^{-\bar y_i[n]} +\bar y_i[n]-1\right)\geq \sum_{k\neq i}\hb_{ki}^H\Sb_k\hb_{ki} + \sigma_i^2,$ and thus
\ifconfver
\begin{align*}
    &\!\!\!\!U(\Sb_1,\ldots, \Sb_K~|~ \{\bar y_i[n]\}_{i=1}^K) \notag \\
    &\leq\log_2\left(1+\frac{\hb_{ii}^H\Sb_i\hb_{ii}}{\sum_{k\neq i}\hb_{ki}^H\Sb_k\hb_{ki}+\sigma_i^2}\right)
     \triangleq U(\Sb_1,\ldots, \Sb_K),
\end{align*}
\else
\begin{align*}
    U(\Sb_1,\ldots, \Sb_K~|~ \{\bar y_i[n]\}_{i=1}^K) &\leq \log_2\left(1+\frac{\hb_{ii}^H\Sb_i\hb_{ii}}{\sum_{k\neq i}\hb_{ki}^H\Sb_k\hb_{ki}
    + \sigma_i^2}\right) \\
    &\triangleq U(\Sb_1,\ldots, \Sb_K),
\end{align*}
\fi
i.e., $U(\Sb_1,\ldots, \Sb_K\,|\,\{\bar y_i[n]\}_{i=1}^K)$ is a universal lower bound of the original objective function $U(\Sb_1,\ldots, \Sb_K)$.
In addition, one can verify that $U(\Sb_1,\ldots, \Sb_K\,|\, \{\bar y_i[n]\}_{i=1}^K)$ and its gradient are locally tight, i.e.,
\ifconfver
\begin{align*}
  &U(\Sb_1^\star[n-1],\ldots, \Sb_K^\star[n-1]~|~ \{\bar y_i[n]\}_{i=1}^K)\\
 &=U(\Sb_1^\star[n-1],\ldots, \Sb_K^\star[n-1]),\\
  &\frac{\partial U(\Sb_1,\ldots, \Sb_K| \{\bar y_i[n]\}_{i=1}^K) }{\partial\Sb_i}
  \bigg |_{{(\Sb_1,\ldots, \Sb_K)=(\Sb_1^\star[n-1],\ldots, \Sb_K^\star[n-1])}}\\
  &=\frac{\partial U(\Sb_1,\ldots, \Sb_K) }{\partial \Sb_i}
  \bigg |_{{(\Sb_1,\ldots, \Sb_K)=(\Sb_1^\star[n-1],\ldots, \Sb_K^\star[n-1]).}}
\end{align*}
\else
\begin{align*}
  &U(\Sb_1^\star[n-1],\ldots, \Sb_K^\star[n-1]~|~ \{\bar y_i[n]\}_{i=1}^K)
  =U(\Sb_1^\star[n-1],\ldots, \Sb_K^\star[n-1]), \\
  &\frac{\partial U(\Sb_1,\ldots, \Sb_K~|~ \{\bar y_i[n]\}_{i=1}^K) }{\partial \Sb_i}
  \bigg |_{\substack{(\Sb_1,\ldots, \Sb_K)\\=(\Sb_1^\star[n-1],\ldots, \Sb_K^\star[n-1])}}
  =\frac{\partial U(\Sb_1,\ldots, \Sb_K) }{\partial \Sb_i}
  \bigg |_{\substack{(\Sb_1,\ldots, \Sb_K)\\=(\Sb_1^\star[n-1],\ldots, \Sb_K^\star[n-1]).}}
\end{align*}
\fi
Therefore, Algorithm 1 in essence is a SUM method in \cite{RazaviHongLuo2012}. According to \cite[Algorithm 1]{RazaviHongLuo2012}, any limit point generated by the SUM algorithm is a stationary point of the original problem. Proposition \ref{proposition: conv of sca} is thus proved.
\hfill{$\blacksquare$}

\footnotesize\vspace{-3.15pt}
\bibliography{Ref_eharvest}

\begin{thebibliography}{10}
\providecommand{\url}[1]{#1}
\csname url@samestyle\endcsname
\providecommand{\newblock}{\relax}
\providecommand{\bibinfo}[2]{#2}
\providecommand{\BIBentrySTDinterwordspacing}{\spaceskip=0pt\relax}
\providecommand{\BIBentryALTinterwordstretchfactor}{4}
\providecommand{\BIBentryALTinterwordspacing}{\spaceskip=\fontdimen2\font plus
\BIBentryALTinterwordstretchfactor\fontdimen3\font minus
  \fontdimen4\font\relax}
\providecommand{\BIBforeignlanguage}[2]{{%
\expandafter\ifx\csname l@#1\endcsname\relax
\typeout{** WARNING: IEEEtran.bst: No hyphenation pattern has been}%
\typeout{** loaded for the language `#1'. Using the pattern for}%
\typeout{** the default language instead.}%
\else
\language=\csname l@#1\endcsname
\fi
#2}}
\providecommand{\BIBdecl}{\relax}
\BIBdecl

\bibitem{ChaoGlobecom2012}
C.~Shen, W.-C. Li, and T.-H. Chang, ``Simultaneous information and energy
  transfer: {A} two-user {MISO} interference channel case,'' in \emph{Proc.
  IEEE GLOBECOM}, Anaheim, USA, Dec. 3-7 2012.

\bibitem{Ozel2011}
O.~Ozel, K.~Tutuncuoglu, J.~Yang, S.~Ulukus, and A.~Yener, ``Transmission with
  energy harvesting nodes in fading wireless channels: {Optimal} policies,''
  \emph{IEEE J. Sel. Areas Commun.}, vol.~29, no.~8, pp. 1732--1743, Sept.
  2011.

\bibitem{Xu2012}
J.~Xu and R.~Zhang, ``Throughput optimal policies for energy harvesting
  wireless transmitters with non-ideal circuit power,'' \emph{IEEE J. Sel.
  Areas Commun.}, vol.~32, no.~12, pp. 1--11, Dec. 2012.

\bibitem{Huang_Zhang_Cui2012}
C.~Huang, R.~Zhang, and S.~Cui, ``Throughput maximization for the {Gaussian}
  relay channel with energy harvesting constraints,'' \emph{IEEE J. Sel. Areas
  Commun.}, vol.~31, no.~8, pp. 1--11, Aug. 2013.

\bibitem{Tutuncuoglu2012_JCN}
K.~Tutuncuoglu and A.~Yener, ``Sum-rate optimal power policies for energy
  harvesting transmitters in an interference channel,'' \emph{J. Commun.
  Netw.}, vol.~14, no.~2, pp. 151--161, April 2012.

\bibitem{Lee2012}
S.~Lee, K.~Huang, and R.~Zhang, ``Cognitive energy harvesting and transmission
  from a network perspective,'' in \emph{Proc. IEEE ICCS}, Singapore, Nov.
  21-23 2012, pp. 225--229.

\bibitem{Kurs2007}
A.~Kurs, A.~Karalis, R.~Moffatt, J.~D. Joannopoulos, P.~Fisher, and
  M.~Solja\v{c}i\'{c}, ``Wireless power transfer via strongly coupled magnetic
  resonances,'' \emph{Science}, vol. 317, no. 5834, pp. 83--86, July 2007.

\bibitem{Karalis2008}
A.~Karalis, J.~Joannopoulos, and M.~Solja\v{c}i\'{c}, ``Efficient wireless
  non-radiative mid-range energy transfer,'' \emph{Annals of Physics}, vol.
  323, no.~1, pp. 34--48, Jan. 2008.

\bibitem{Dolgov2010}
A.~Dolgov, R.~Zane, and Z.~Popovic, ``Power management system for online low
  power {RF} energy harvesting optimization,'' \emph{IEEE Trans. Circuits and
  Syst. I}, vol.~57, no.~7, pp. 1802--1811, July 2010.

\bibitem{Nintanavongsa2012}
P.~Nintanavongsa, U.~Muncuk, D.~Lewis, and K.~Chowdhury, ``Design optimization
  and implementation for {RF} energy harvesting circuits,'' \emph{IEEE J.
  Emerg. and Sel. Topics in Circuits and Syst.}, vol.~2, no.~1, pp. 24--33,
  Mar. 2012.

\bibitem{Varshney2008}
L.~R. Varshney, ``Transporting information and energy simultaneously,'' in
  \emph{Proc. IEEE ISIT}, Toronto, Canada, July 6-11 2008, pp. 1612--1616.

\bibitem{Grover2010}
P.~Grover and A.~Sahai, ``Shannon meets {Tesla}: {Wireless} information and
  power transfer,'' in \emph{Proc. IEEE ISIT}, Austin, USA, June 13-18 2010,
  pp. 2363--2367.

\bibitem{Fouladgar2012}
A.~M. Fouladgar and O.~Simeone, ``On the transfer of information and energy in
  multi-user systems,'' \emph{IEEE Commun. Lett.}, vol.~16, no.~11, pp.
  1733--1736, Nov. 2012.

\bibitem{Zhou2012}
X.~Zhou, R.~Zhang, and C.~K. Ho, ``Wireless information and power transfer:
  {Architecture} design and rate-energy tradeoff,'' in \emph{Proc. IEEE
  GLOBECOM}, Anaheim, USA, 2012, pp. 3982--3987.

\bibitem{Zhou2012_jrnl}
------, ``Wireless information and power transfer: {Architecture} design and
  rate-energy tradeoff,'' \emph{ArXiv e-prints}, pp. 1--30, May 2012.

\bibitem{Zhangrui2011}
R.~Zhang and C.~K. Ho, ``{MIMO} broadcasting for simultaneous wireless
  information and power transfer,'' \emph{IEEE Trans. Wireless Commun.},
  vol.~12, no.~5, pp. 1989--2001, May 2013.

\bibitem{Shi2013}
Q.~Shi, W.~Xu, and D.~Li, ``Joint beamforming and power splitting for
  multi-user {MISO} {SWIPT} system,'' \emph{ArXiv e-prints}, pp. 1--6, Mar.
  2013.

\bibitem{WCLi2013}
W.-C. Li, T.-H. Chang, C.~Lin, and C.-Y. Chi, ``Coordinated beamforming for
  multiuser {MISO} interference channel under rate outage constraints,''
  \emph{IEEE Trans. Signal Process.}, vol.~61, no.~5, pp. 1087--1103, March
  2013.

\bibitem{Jorswieck08}
E.~Jorswieck, E.~Larsson, and D.~Danev, ``Complete characterization of the
  {Pareto} boundary for the {MISO} interference channel,'' \emph{IEEE Trans.
  Signal Process.}, vol.~56, no.~10, pp. 5292--5296, Oct. 2008.

\bibitem{Lindblom2011}
J.~Lindblom, E.~Karipidis, and E.~G. Larsson, ``Closed-form parameterization of
  the {Pareto} boundary for the two-user {MISO} interference channel,'' in
  \emph{Proc. IEEE ICASSP}, Prague, Czech, May 22-27 2011, pp. 3372--3375.

\bibitem{Liu2012_IFC}
L.~Liu, R.~Zhang, and S.~Lambotharan, ``Achieving global optimality for
  weighted sum-rate maximization in the {K}-user {Gaussian} interference
  channel with multiple antennas,'' \emph{IEEE Trans. Wireless Commun.},
  vol.~11, no.~5, pp. 1933--1945, May 2012.

\bibitem{ZhangCui2010}
R.~Zhang and S.~Cui, ``Cooperative interference management with {MISO}
  beamforming,'' \emph{IEEE Trans. Signal Process.}, vol.~58, no.~10, pp.
  5450--5458, Oct. 2010.

\bibitem{Shang2011}
X.~Shang, B.~Chen, and H.~V. Poor, ``Multi-user {MISO} interference channels
  with single-user detection: {Optimality} of beamforming and the achievable
  rate region,'' \emph{IEEE Trans. Inf. Theory}, vol.~57, no.~7, pp.
  4255--4273, July 2011.

\bibitem{Luo2008}
Z.-Q. Luo and S.~Zhang, ``Dynamic spectrum management: {Complexity} and
  duality,'' \emph{IEEE J. Sel. Topics Signal Process.}, vol.~2, no.~1, pp.
  57--73, Feb. 2008.

\bibitem{Marks1978}
B.~R. Marks and G.~P. Wright, ``A general inner approximation algorithm for
  nonconvex mathematical programs,'' \emph{Operations Research}, vol.~26, pp.
  681--683, 1978.

\bibitem{cvx}
M.~Grant and S.~Boyd, ``{CVX}: Matlab software for disciplined convex
  programming, version 1.22,'' \url{http://cvxr.com/cvx}, Aug. 2012.

\bibitem{RazaviHongLuo2012}
M.~Razaviyayn, M.~Hong, and Z.~Luo, ``A unified convergence analysis of block
  successive minimization methods for nonsmooth optimization,'' \emph{SIAM J.
  Optimization}, vol.~23, no.~2, pp. 1126--1153, 2013.

\bibitem{ChaoTSP2012}
C.~Shen, T.-H. Chang, K.-Y. Wang, Z.~Qiu, and C.-Y. Chi, ``Distributed robust
  multicell coordianted beamforming with imperfect {CSI}: {An ADMM} approach,''
  \emph{IEEE Trans. Signal Process.}, vol.~60, no.~6, pp. 2988--3003, 2012.

\bibitem{Jafar2008}
V.~R. Cadambe and S.~A. Jafar, ``Interference alignment and degrees of freedom
  of the {K}-user interference channel,'' \emph{IEEE Trans. Inf. Theory},
  vol.~54, pp. 3425--3441, Aug. 2008.

\bibitem{Huang2010}
Y.~Huang and D.~Palomar, ``Rank-constrained separable semidefinite programming
  with applications to optimal beamforming,'' \emph{IEEE Trans. Signal
  Process.}, vol.~58, no.~2, pp. 664--678, Feb. 2010.

\bibitem{BK:BoydV04}
S.~Boyd and L.~Vandenberghe, \emph{Convex {O}ptimization}.\hskip 1em plus 0.5em
  minus 0.4em\relax Cambridge, UK: Cambridge University Press, 2004.

\bibitem{Charnes1962}
A.~Charnes and W.~W. Cooper, ``Programming with linear fractional functions,''
  \emph{Naval Res. Logist. Quarter.}, vol.~9, pp. 181--186, Dec. 1962.

\end{thebibliography}
\end{document}